\documentclass{aa}

\usepackage{graphicx}
\usepackage{txfonts}
\usepackage{natbib}
\usepackage[dvipsnames,usenames,table]{xcolor}
\usepackage[normalem]{ulem}
\usepackage{longtable}
\usepackage{longfigure}

\usepackage{txfonts}
\begin{document} 

\title{A population analysis of Galactic Miras with light-curve asymmetries\thanks{Table of sample stars is only available in electronic form at the CDS via anonymous ftp to cdsarc.u-strasbg.fr (130.79.128.5) or via http://cdsweb.u-strasbg.fr/cgi-bin/qcat?J/A+A/.}}

\author{
S. Uttenthaler\inst{\ref{inst_iap}}
\and
P. Merchan-Benitez\inst{\ref{inst_extrema1}}}

\institute{
Institute of Applied Physics, TU Wien, Wiedner Hauptstra\ss e 8-10, 1040 Vienna, Austria\label{inst_iap};\\ \email{stefan.uttenthaler@gmail.com}
\and
Faculty of Science, University of Extremadura, 06011 Badajoz, Spain\label{inst_extrema1}; \email{pedromer@hotmail.com}
}

\date{Received September 15, 1996; accepted March 16, 1997}

\abstract 
{In a recent publication, we established a close relationship between light-curve asymmetries in Mira variables and indicators of their dust mass-loss rate. The light-curve asymmetries appear to be related to the stars' third dredge-up (3DUP) activity.}
{We aim to reveal the evolutionary status of M-type Miras with light-curve asymmetries ('bumps') within the spectral sequence M -- S -- C, determine their mass-loss properties, and check possible evolutionary scenarios.}
{We analysed a sample of 3100 Miras collected from the ASAS database, distinguishing between symmetric and asymmetric light curves. We determined their periods, luminosity functions, and period-luminosity relations, their locations relative to the Galactic midplane, as well as mass-loss rate indicators through the 2MASS-WISE colours and the {\it Gaia}-2MASS diagram.}
{The M-type Miras with symmetric light curves are generally found to have shorter periods, lower luminosities, a larger average distance to the Galactic midplane, and lower initial masses than the M-type Miras with asymmetries. In addition, 25 Miras are proposed as candidates for new carbon stars.}
{We propose that the M-type Miras have two distinct populations: M-type Miras with symmetric light curves, which have lower initial mass progenitors than M-type Miras with asymmetries, which show signs of 3DUP activity and are the link to the S-type Miras.}


\titlerunning{Miras with meandering periods}
\authorrunning{Uttenthaler \& Merchan-Benitez}
\maketitle

\section{Introduction} \label{sec:Introduction}

Mira stars are long-period variables (LPVs) in the asymptotic giant branch (AGB) phase of stellar evolution, characterised by large amplitudes of variation \citep[$\Delta V>2\fm5$,][]{Samus'2017} and pulsation periods of $P\sim100-1000$\,d. In contrast to other LPV classes, Miras frequently also show emission lines of hydrogen and primarily neutral metals around their maximum visual light that arise from shock waves in the atmospheres \citep{Merrill1952,Willson1976}. During the thermally pulsing (TP-)AGB phase, Mira variables produce dust grains efficiently, have high mass-loss rates \citep[$\dot{M} \sim 10^{-8} - 10^{-4}$ $M_{\sun}$ $yr^{-1}$,][]{Hofner2018}, and enrich the interstellar medium in dust and products of nucleosynthesis. These products are brought to the stellar surface by a deep mixing event called the third dredge-up (3DUP). In a recent study, \citet[hereafter Paper~I]{Merchan-Benitez2023} selected a sample of 548 Miras in the solar neighbourhood to study both the time evolution of their periods and the relation between the shape of their light curves and their mass-loss properties. 

The semi-regular (SR) and irregular (L) variables that populate both the red giant branch (RGB) and the AGB also belong to the LPVs, but are distinguished from the Mira variables by their smaller amplitudes ($\Delta V<2\fm5$), less regular light curves, and the general absence of emission lines. However, these criteria have been shown to be arbitrary, so that there is a smooth transition from Mira variables to SRV and L variables \citep[e.g.][]{Bedding1998,Lebzelter2009,Trabucchi2021}. Some authors have included the L variables in the SRV class, arguing that their irregularity is simply a result of too few observations \citep[e.g.][]{Lebzelter1995,Kerschbaum1999}.

About a century ago, \citet{Campbell1925} and \citet{Ludendorff1928} established classification systems for light-curve shapes of Mira variables. \citet{Ludendorff1928} separated them into three main groups: $\alpha$, $\beta$, and $\gamma$. \citet{Campbell1955} introduced an asymmetry factor in the classification, defined as the rise time from minimum to maximum light relative to the mean pulsation period, later also used by \citet{Vardya1988} in their work. \citet{Lebzelter2011} used the sum of the squared differences between the light curve of a star and a sinusoidal reference curve ($\chi^2$) and found that $\sim30\%$ of the sample stars have light curves that significantly deviate from a purely sinusoidal shape. In addition, a connection between atmospheric chemistry and the shape of the light curve was found, with a higher fraction of S and C stars having non-sinusoidal variations compared to M stars, although the conclusions were limited by the small number of S and C Miras in the sample. In Paper~I, we proposed a division into two groups: Group~A (Asymmetric), with Miras showing 'bumps' or abrupt changes in the slope of the ascending branch of their light curves or double maxima, and Group~S (Symmetric), with symmetric, nearly sinusoidal light curves. It was found that $\sim$ 37\% of the Miras in the sample had asymmetries in their light curves. The fractions of asymmetric light curves were 31\%, 75\%, and 80\% among the M-, S-, and C-type stars, respectively, which confirms the results of \citet{Lebzelter2011}. More recently, \citet{Hoai2025} studied in depth a more limited sample of well-observed Miras and the relation of light-curve shape parameters to the stellar evolutionary state, confirming the relations with spectral types.

Several works established a relation between pulsation period and light-curve asymmetries on the one hand and mass-loss indicators on the other. \citet{Bowers1975} and \citet{Bowers1977} found a relationship between the slope of the rising branch of the light curves and OH emission at millimetre wavelengths, at least for stars with $300\lesssim P\lesssim450$\,d, with OH and non-OH Miras reflecting different pulsation characteristics and, hence, mass-loss rates. Recently, \citet{Smith2023} established that the conditions for OH emission to occur include the colour $J-K_{\rm S} \gtrsim 1.4$ and a period generally longer than 316 days. In addition, Miras with asymmetric light curves have silicate emission features at 9.7 and 20\,$\mu$m, while those with more symmetric light curves show much weaker emission features at 12 and 20\,$\mu$m \citep{Vardya1986,Onaka1989}.
On the other hand, there is a high probability of H$_{2}$O detection for symmetric light curves, it being almost null for asymmetric light curves \citep{Vardya1987}. Finally, \citet{LeBertre1992} and \citet{Winters1994} explored a possible relation between light-curve shape and dust formation.

In Paper~I, a close relationship between asymmetries in the light curves and a possible post-3DUP evolutionary state was discovered. M-type Miras with light-curve bumps (hereafter referred to as M(B) Miras) are located in the same area in a $K-[22]$ diagram where Tc-rich, post-3DUP stars are preferentially located, while most M-type Miras without bumps (hereafter, M(NB)-type Miras) were located in the same area where Tc-poor stars are preferentially located. In addition, the presence of Tc in many of the M(B)-type Miras and high $^{12}$C/$^{13}$C isotope ratios support the hypothesis that they are in the TP-AGB phase.

The main objective of the present article is to follow up on the differences between the M(B) and M(NB) Miras and related LPVs on a population level. We have collected a large sample of Galactic LPVs with well-observed light curves to distinguish those with bumps from those without and characterise them in more detail. We have also analysed indicators of progenitor mass and evolutionary stage to see how they fit in the well-established M -- S -- C evolutionary sequence.

\section{Sample stars and data reduction}\label{sec:Sample stars and data reduction}

\subsection{Collecting the sample and supplementary data}\label{sec:samplecollection}

For the present paper, we require photometric data for a large number of LPVs that cover at least three to four pulsation cycles. Therefore, we have extended the sample of Paper~I by data collected in the ASAS \citep[All Sky Automated Survey,][]{Pojmanski2005} and ASAS-SN \citep[All-Sky Automated Survey for Supernovae,][]{Shappee2014} Catalogue of Variable Stars III \citep{Jayasinghe2019}. The ASAS light curves from the ASAS-1 through ASAS-3 project phases cover the time interval from April 1997 to December 2010, while the ASAS-SN $V$-band observations were made between 2013 and 2018. The $V$-band photometric data observed by ASAS typically cover five to ten pulsation cycles, have a sufficiently high number of observations, and have a limiting magnitude of about 14\fm5. However, as Miras are characterised by large pulsation amplitudes, this magnitude limit means that the light minima of fainter Miras can be below the detection threshold, making it difficult to identify asymmetries. Thus, we used the ASAS-SN database\footnote{\url{http://asas-sn.osu.edu/variables}} to supplement the ASAS light curves where necessary, as it includes photometric data up to 18\fm5. Although ASAS-SN covers fewer pulsation cycles than ASAS, it allows us to complete the phased ASAS light curves in their minima in many cases. Therefore, the combined use of both databases allows us to reach the limits required for our objective, both in the number of cycles covered and in the depth of the exposures.

We downloaded the data and light curves of all variables classified as Miras in the ASAS catalogue from the project website\footnote{\url{https://www.astrouw.edu.pl/asas/?page=acvs}}. This yielded a total of 2901 stars. However, according to \citet{Pojmanski2002}, ASAS did not apply the classical Mira amplitude criterion, but rather all variables with $\Delta V>2\fm0$ and $P>70$\,d were classified as Miras. Some SRVs, L variables, and even cataclysmic variables might have also been classified as Miras. Thus, we searched for their counterparts in the most recent version of the General Catalogue of Variable Stars \citep[GCVS,][]{Samus'2017} and in SIMBAD \citep{Wenger2000} to determine their variability type. Table~\ref{Tab:sample} summarises the results of this cross-match and how the stars distribute on the different chemical spectral types. We see from this that there are a total of 181 SRVs and 13 L variables according to GCVS in the sample. 

\begin{figure}[!t]
\includegraphics[width=\linewidth]{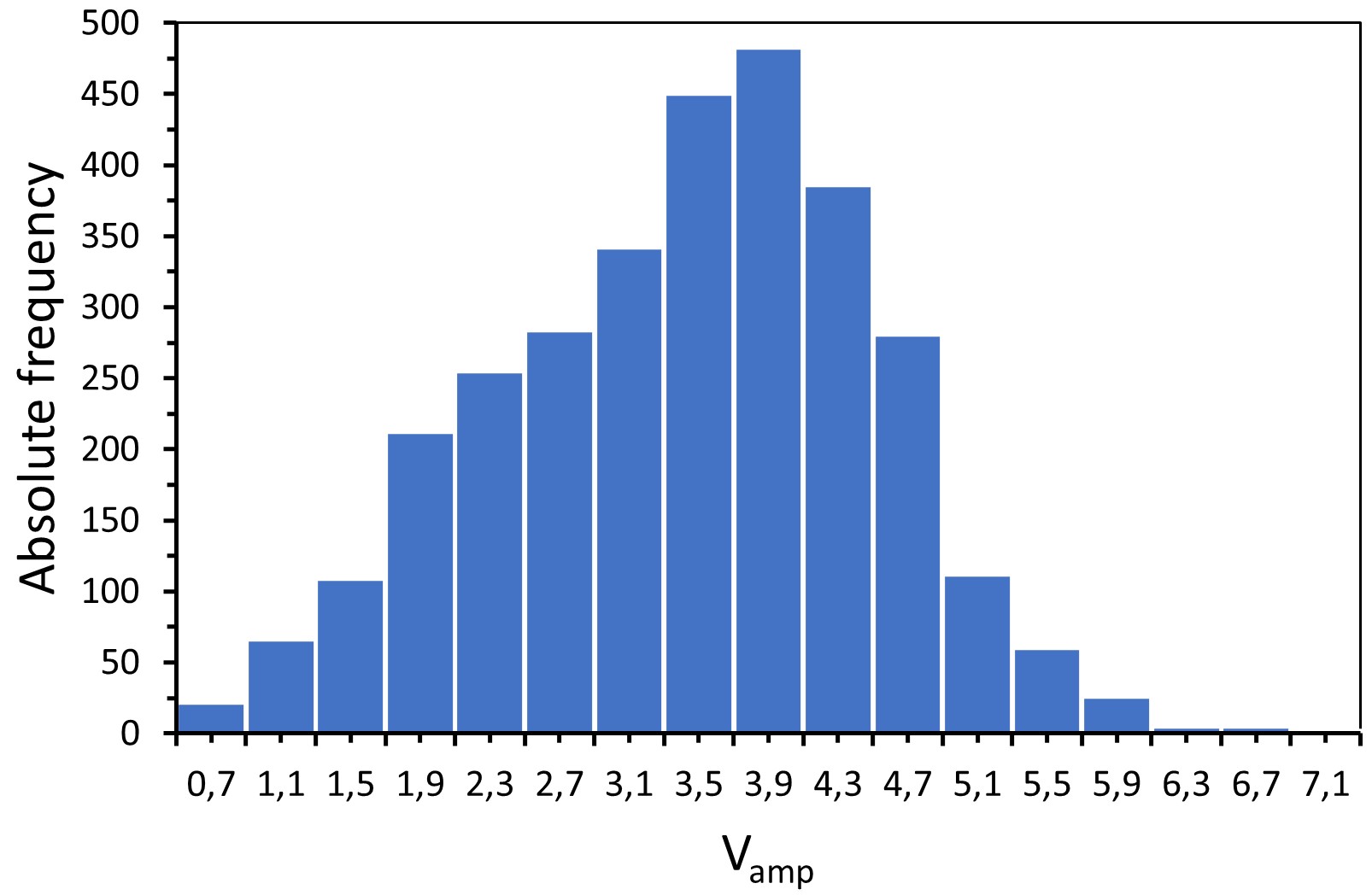}
\centering
\caption{Histogram of $V$-band amplitudes for the 3100 LPVs in the sample, with bins of 0.4 mag. 
}
\label{Fig:Vamp Histogram}
\end{figure}

We added to this the 227 Miras from Paper~I that were not already included in the ASAS database, yielding 3128 stars. Of these, 28 stars were discarded for reasons detailed in Table~\ref{Tab:Discarded stars} in the appendix. The final sample consists of 3100 LPVs. Their distribution on variability and spectral type is detailed in Table~\ref{Tab:sample}. The histogram of $V$-band amplitudes for these variables, compiled from the ASAS database, is shown in Fig.~\ref{Fig:Vamp Histogram}, where it can be seen that there is a non-negligible fraction of $664/3100\approx21.4\%$ with amplitudes $\Delta V<2\fm5$ that have been included in this project. However, as we show in Sect.~\ref{subsect: P-L diagrms}, almost all of the stars closely follow the period-luminosity sequence formed by the Miras, so we can be confident that the overwhelming majority of our sample LPVs are fundamental-mode, Mira-like variable stars. For simplicity, we will thus refer to the sample stars as Mira-like variables or Miras throughout the paper. All stars will be analysed independently of their previous classification.

\begin{table}
\caption{Variability and spectral types in the sample according to the GCVS.}
\label{Tab:sample}
\centering
\begin{tabular}{cccccc}
\hline\hline
Var.\ type    &      & SpT &     &         & Total\\
              &    M &   S &   C & Indef.\ &      \\ 
\hline
Mira          & 1433 & 107 & 132 &    1234 & 2906 \\
Semi-regular* &   73 &   2 &   7 &      99 &  181 \\
Irregular     &    7 &   1 & --- &       5 &   13 \\
Total         & 1513 & 110 & 139 &    1338 & 3100 \\
\hline
\end{tabular}
\tablefoot{* Semi-regulars of the SRc and SRd types consisting of red supergiants and yellow giants have not been considered in the sample. Only one SRd type, namely W~LMi, has been retained because of its ambiguous spectral classification, G2e-K2e(M3).}
\end{table}

\begin{figure*}[!t]
\includegraphics[scale=0.65]{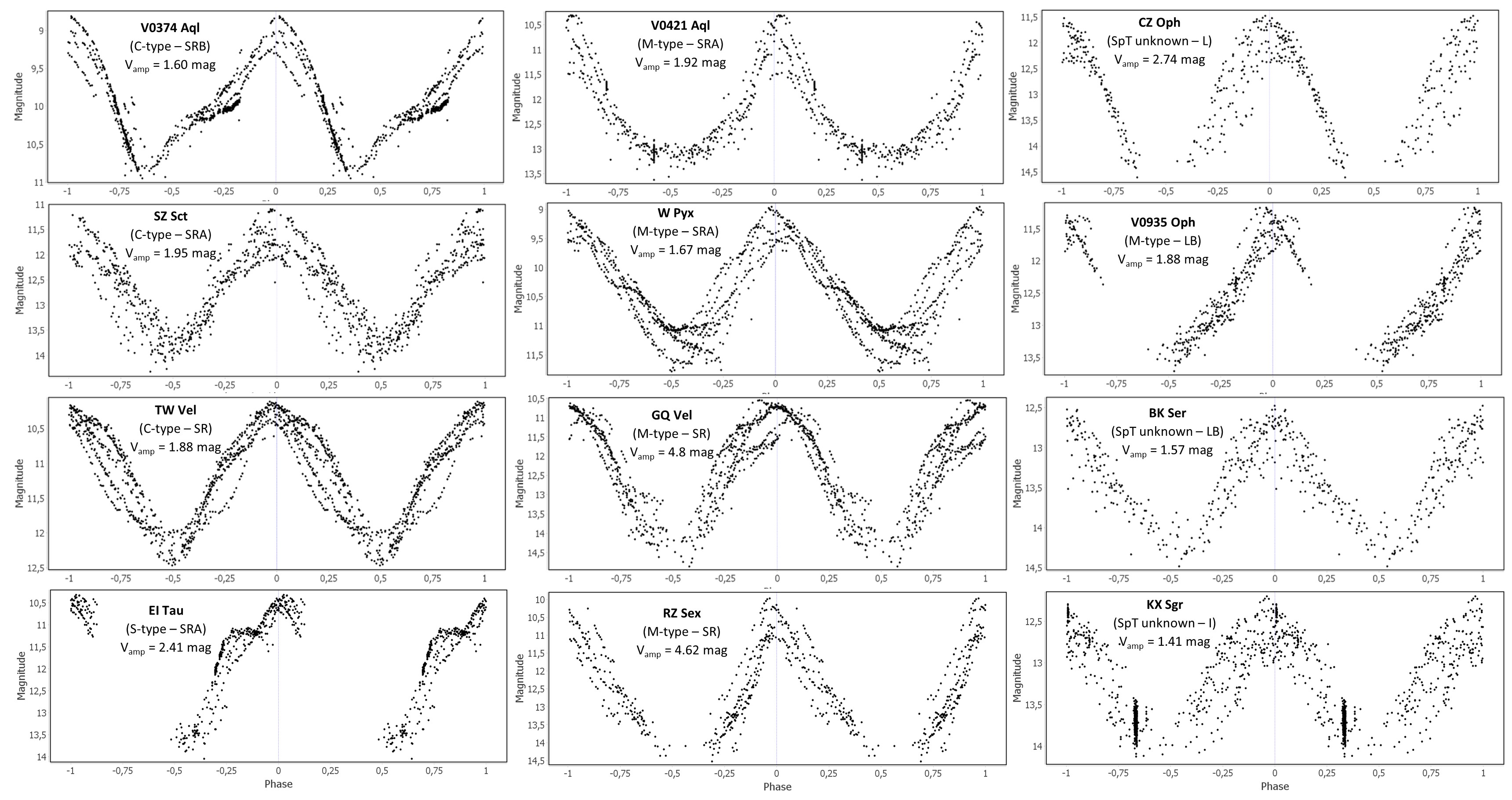}
\centering
\caption{Phase diagrams of stars with a previous GCVS classification as SRV and irregular variables, with different spectral types and amplitude ranges (see text). The data demonstrate they are Mira-like variables.
}
\label{Fig:SRVs}
\end{figure*}

We determined the pulsation period for all sample stars using the Fourier routine DC-DFT in the AAVSO software package VStar, except for the 548 Miras included in Paper~I, for which we adopted the mean period calculated there. For SRVs and L variables according to GCVS (Table~\ref{Tab:sample}), we analysed their phase diagram using the period previously obtained by FFT. Figure~\ref{Fig:SRVs} shows several examples of these phase diagrams. Comparing the phase diagrams with those of genuine Miras suggests that the light curves are very regular and actually similar to those of Miras, albeit with amplitudes below the classical Mira criterion ($\Delta V<2\fm5$). They also resemble the SRas of \citet[][their Fig.~4d]{2001ApJ...552..289A} but, unlike most of the SRVs in that work, only two of our sample stars have periods slightly shorter than 100\,d. Several S- and C-type stars are included in the left-hand panels and several M-type stars in the centre panels, all with a prior GCVS classification as SRVs. In the right-hand panels, four more stars are included with a previous GCVS classification as irregulars, some with unknown spectral types.

In some of the light curves shown in Fig.~\ref{Fig:SRVs}, the presence of possible bumps can be clearly seen (e.g. V0374~Aql, EI~Tau, or GQ~Vel), which may be the reason why they have been previously classified as SR or L variables. Therefore, we will not discriminate on the basis of the amplitudes and irregularities of their light curves, eliminating only those that do not show signs of periodicity in their phase diagrams, as is the case for 20 of the 28 excluded variables (Table~\ref{Tab:Discarded stars}). In Fig.~\ref{Fig:SRVs}, we can also notice that some stars are below the faint limit at minimum light, as in RZ~Sex and CZ~Oph, so the amplitude may be underestimated, and most likely they are normal Miras that are misclassified in ASAS because they are below the detection limit at minimum light. In addition, a related problem may arise for stars with large annual gaps, as in EI~Tau and V0935~Oph, which could also complicate the classification of the variability type.

For each object in the sample, we cross-identified its counterparts in the Two-Micron All-Sky Survey \citep[2MASS,][]{Skrutskie2006}, Wide-field Infrared Survey Explorer \citep[WISE,][]{Wright2010}, and {\it Gaia} Early Data Release 3 \citep[{\it Gaia} DR3,][]{Gaia2022} by matching it to the closest object in the ASAS catalogue. In 2MASS, only photometric data with a signal-to-noise ratio (S/N) $>5$ (quality flags A, B, and C) were considered, while in the AllWISE programme \citep{Cutri2021}, only unsaturated data with quality flags A, B, and C were used. For the saturated stars, we checked the flux provided by the unWISE mission \citep{Schlafly2019}, in which the quality factor of the data is defined between 0 (saturated detections) and 1 (good detections). We found for all of them that this quality factor never exceeds 0.1, so they have not been taken into account in the sections in which the WISE colours are used.

The spectral types of the stars were collected in the following way. For the 548 variables in Paper~I, we took the information collected there. In order of priority, we used the following sources of spectral type classification of the remaining stars: the GCVS \citep{Samus'2017}, SIMBAD \citep{Wenger2000}, the Catalogue of Stellar Spectral Classifications by \citet{Skiff2014}, the LAMOST DR4 database \citep{Luo2018}, and the AGB star catalogue of \citet{Shu2022}. No additional matches were found in the General Catalog of Galactic Carbon Stars \citep{Alksnis2001} or in the General Catalog of S Stars \citep{Stephenson1984}. For the identification in all these catalogues, we considered objects with a valid cross-reference to the coordinates listed in the 2MASS PSC catalogue with a distance of less than 2\arcsec. The distribution of the sample stars on the spectral types is listed in columns~2 to 5 of Table~\ref{Tab:sample}.

\subsection{Classification in terms of bumps and asymmetries}\label{subsect:Classification in terms of bumps and asymmetries}

We carefully analysed the light curves of the 2552 new sample stars, following the same criteria as Paper~I and separating them into the two groups introduced there (see the examples in Fig.~8 of Paper~I for the definition of Mira variables with and without bumps):

Group~A (Asymmetric): Stars with apparent asymmetries in their light curves, such as bumps on the rising branch, double maxima, or abrupt changes in the slope of the rising branch. Given the similarity of all these anomalies, the stars form a relatively homogeneous group, as the asymmetries may correspond to a larger- or smaller-scale expression of similar phenomena in the atmospheres of stars.

Group~S (Symmetric): Stars without obvious asymmetries in the ascending branch of their light curves, including also in this group doubtful cases in which the possible observed anomalies occur only in a few cycles or are so weak that we cannot state with certainty their existence.

We expect Group~A to be relatively pure as a star will be classified in Group~A only if it has obvious anomalies. In contrast, Group~S may contain some stars that would be classified as Group~A if better data were available. This aspect may introduce some uncertainty in our results, but the large number of stars in our selection and the use of two different surveys to inspect the light curves minimise these uncertainties (up to four different surveys for Paper~I variables).

Besides this subjective classification based on visual light curve inspection that may not be fully reproducible and lead to misclassifications, we also made efforts to develop objective criteria based on numerical methods to classify the stars. Unfortunately, these efforts were ultimately unsuccessful. We adopted a sub-sample of 157 stars from Paper~I with well-observed light curves from different photometric databases and tried to reproduce the visual bump classification with different numerical methods. We focused on the $\chi$ parameters of \citet{Lebzelter2011}: $\chi^2$, the squared difference between the normalised average light curve and a sinusoidal reference light curve; $\chi_1$, the total difference between the observed light curve and the reference light curve on the rising branch; and $\chi_2$, the same as $\chi_1$ for the descending branch of the light curve. As in \citet{Lebzelter2011}, these values were calculated from 20 points equally distributed in phase over the normalised averaged light curve (see their Sect.~3 for the exact definitions). In the presence of bumps, one may expect $\chi_1$ to have a relatively large (absolute) value, whereas the value of $\chi_{2}$ should not depend on the presence of bumps. The difference $\chi_{1}-\chi_{2}$ might also have an enhanced value for stars with bumps. In addition, bumps may lead to a relatively sharply rising light curve. Therefore, the maximum slope between two adjacent data points in the normalised averaged light curve was calculated for each star. In addition to the $V$ band, data from the $I$ band were also analysed where available, for example, from the KWS survey \citep{Maehara2014}\footnote{\url{http://kws.cetus-net.org/~maehara/VSdata.py}}. The distribution of the stars with and without bumps, respectively, indeed differed slightly and in the expected direction. However, the two groups did not separate into two sufficiently pure groups in any of the numerical parameters or combinations thereof to confidently replace the visual inspection. Possible reasons are observational uncertainties, cycle-to-cycle variations of the light curves, and a possible phase shift of the sinusoidal reference curve induced by the presence of light curve bumps that could impact the $\chi_{1}-\chi_{2}$ parameter. We, therefore, continued with the results of the visual light curve inspection for the remainder of the work.

Among the sample stars with a known spectral type (1762 stars), we classified 596 Miras in Group~A, which represents a fraction of 596/1762 $\approx33.8$\%. This fraction is in good agreement with $\sim37$\% found in Paper~I and with the fraction of $\sim30$\% of stars with significant deviation from the sinusoidal reference shape found by \citet{Lebzelter2011}. Group~A Miras do not evenly distribute on the chemical spectral types: $84/110\approx76.4$\% of the S-type, $106/139\approx75.5$\% of the C-type, and only $406/1513\approx27.0$\% of the M-type Miras have asymmetries, also in good agreement with Paper~I. Appendix~\ref{Appendix A} includes the analysis for Miras with an unknown spectral type. 

\begin{figure}[!t]
\includegraphics[width=\linewidth]{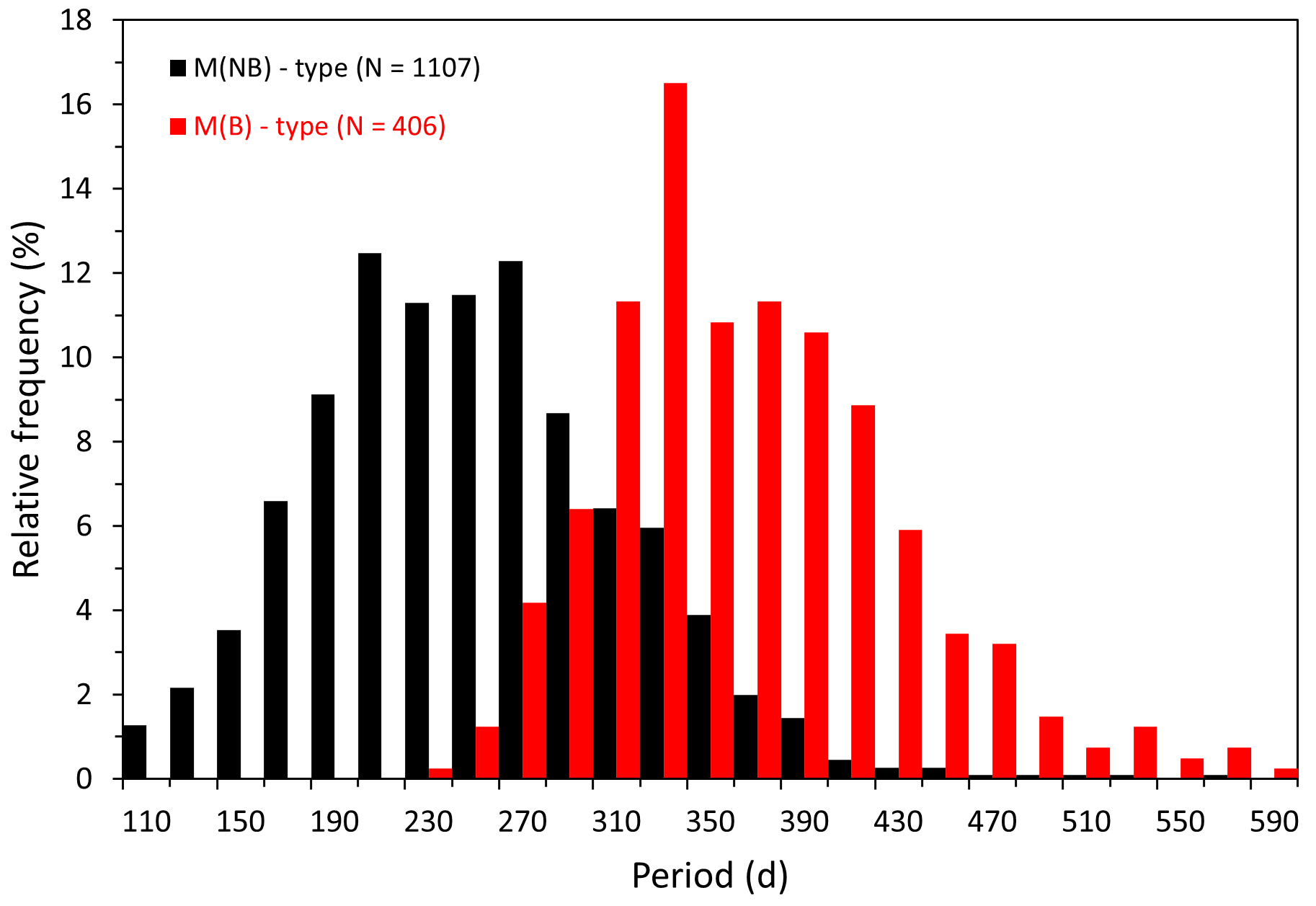}
\centering
\caption{Histogram of the pulsation periods for the M-type Miras in the sample, with bins of 40\,d. See legend to find the number of M(NB)- and M(B)-type Miras.}
\label{Fig:Histogram M - type}
\end{figure}

M-type Miras with and without light-curve asymmetries, respectively, show a clear difference in the period distribution. While M(B)-type Miras have their maximum frequency at $P\approx330$\,d and generally have pulsation periods $P\gtrsim250$\,d, the M(NB)-type Miras reach their maximum at $P\approx220$\,d, as can be appreciated in the histogram in Fig.~\ref{Fig:Histogram M - type}. Considering that the pulsation period is a function of the initial stellar mass \citep[e.g.][]{Whitelock2019}, the difference in the distribution of the pulsation periods suggests a population difference in the initial mass of each group.

\section{Luminosities and Galactic distribution}\label{sect: Luminosities and Galactic distribution}

We adopted the photo-geometric distances, $d$, from \citet{Bailer-Jones2021b} to calculate luminosities and, when not possible, the geometric distances. Photo-geometric distances are generally more accurate and precise for stars with poor parallaxes, also exploiting the fact that stars of a given colour have a restricted range of probable absolute magnitudes. They also provide meaningful distances for negative parallaxes through the use of probabilistic inference, and although they do not provide fully accurate distances, the resulting confidence intervals are significant \citep{Bailer-Jones2021a}. Only for R~Hor, W~Hya, and R~Leo, the distances have been calculated from the Hipparcos parallaxes included in the catalogue of \citet{vanLeeuwen2007}.

The re-normalised unit weight error (RUWE) parameter is often used as a statistical indicator of the quality of {\it Gaia} DR3 astrometric data and represents a normalisation of the $\chi^2$ distribution or unit weight error. The typically adopted limit is $RUWE<1.4$ by \citet{Lindegren2018}, although its exclusive use is not recommended \citep{Andriantsaralaza2022}. However, it is quite safe to use parallaxes with $1.4<RUWE<2.0$, and even parallaxes with values $2.0<RUWE<3.0$, despite the probability being higher that the data are somewhat biased \citep{Maiz-Apellaniz2021}. Recently, \citet[][see their Fig.~1]{Messineo2019} has suggested that $RUWE<2.7$ will yield adequate results. We adopted this value for our astrometric data together with a filter on the parallax accuracy of $\varpi/\sigma_{\varpi}>4.0$, thus eliminating stars with too uncertain parallaxes. These filters imply a significant loss of $\sim30\%$ for stars with a known spectral type (up to $\sim64\%$ for the fainter stars with an unknown spectral type), but ensure a good accuracy of the results obtained. The sample stars with known spectral types that satisfy the parallax uncertainty criterion have distances between 150\,pc and 19.4\,kpc, with a median of 1690\,pc and a standard deviation of 1540\,pc. Those without known spectral types have distances between 340\,pc and 20.6\,kpc, with a median of 3150\,pc and a standard deviation of 2200\,pc. Only stars satisfying the parallax uncertainty criterion are displayed in the figures that show the $M_{K,0}$ luminosity in the following sections.

\subsection{Luminosity functions}\label{subsect:LFs}

We constructed the luminosity functions (LFs) of the different groups of stars with a known spectral type from the absolute magnitudes $M_{K,0}$, shown in Fig.~\ref{Fig:/Mk Histogram SpT} (see Appendix~A for the Miras with an unknown spectral type). $M_{K,0}$ was calculated following the equation $M_{K,0} = K_{\rm S} - A_{K,{\rm S}} - 5\cdot\log_{10}(d) + 5$, where $K_{\rm S}$ is the 2MASS $K_{\rm S}$ magnitude, $d$ is the distance to the star from \citet{Bailer-Jones2021a}, and $A_{K,S}$ is the interstellar extinction in the $K$ band. The $A_{K,{\rm S}}$ values were determined from the 3D map of \citet{Gontcharov2017} and the equations given there. The cross-match with the map grid points was done with the software TOPCAT \citep{Taylor2005}. Since the map of \citet{Gontcharov2017} extends only out to 1200\,pc from the sun, extinction was set to this outer boundary for the more distant stars. This procedure yielded an average interstellar extinction in the $K$ band of 0\fm077 for stars with known spectral types, and 0\fm092 for those without known spectral types. During the review process, we became aware of the larger map of \citet{Gontcharov2023}, which extends in 3D to 2\,kpc from the sun. We compared the extinction values from both maps for the locations of our sample stars, adopting the extinction values from the 2D map of \citet{Gontcharov2023} for the stars still outside the 3D map. This comparison showed that, on average, the difference between both maps is smaller than 0\fm01, with a standard deviation of $\sim0\fm03$. In fact, the extinction values from the 2023 map are slightly lower than those from the 2017 map, except for very few of the most distant and strongly extinguished stars. \citet{Gontcharov2023} showed that there are still significant variations in absolute extinction between the different maps presented in the literature. The relatively small differences between the maps found by us and the results of \citet{Gontcharov2023} convinced us that adopting any other extinction map would not significantly impact or improve the overall results of that paper. We therefore decided to retain the correction based on the \citet{Gontcharov2017} map.

In Fig.~\ref{Fig:/Mk Histogram SpT}, the M(B)-, S-, and C-type Miras have similar distributions with peaks at $M_{K,0}\approx-7.8$ and mean absolute magnitudes of $\langle M_{K,0} \rangle = -7.96 \pm{0.66}$, $\langle M_{K,0} \rangle = -7.84 \pm{0.87}$, and $\langle M_{K,0} \rangle = -7.85 \pm{0.65}$, respectively. In the histogram of the M(B)-type Miras, a tail at the bright end is noticeable, which could be intermediate- to high-mass O-rich AGB stars undergoing hot bottom burning \citep[HBB][]{Garcia-Hernandez2013,Lebzelter2018}. In contrast, the LF of the M(NB)-type Miras peaks at $M_{K,0}\approx-7.2$, as would be expected from the fact that they have generally shorter pulsation periods (see Fig.~\ref{Fig:Histogram M - type}). The mean absolute magnitude is $\langle M_{K,0} \rangle = -7.07 \pm{0.71}$, much lower than that of the other groups. We estimate the RGB tip luminosity from the observed de-reddened RGB tip in the LMC \citep[$m_{\rm TRGB}(K_{\rm S})=11\fm94\pm0.04$,][]{2000A&A...359..601C} and the distance modulus to the LMC \citep[$\mu=18\fm477$,][]{2019Natur.567..200P} to be $M_{\rm K,0}=-6\fm54$. This means that some of the M(NB) Miras are actually fainter than the RGB tip. It is of particular relevance that within the M-type stars there is such an evident difference between the LFs of the M(NB)- and M(B)-type Miras, with the latter similar to the S- and C-type Miras, which suggests that the two groups evolve from different progenitors.

\begin{figure}[t]
\includegraphics[width=\linewidth]{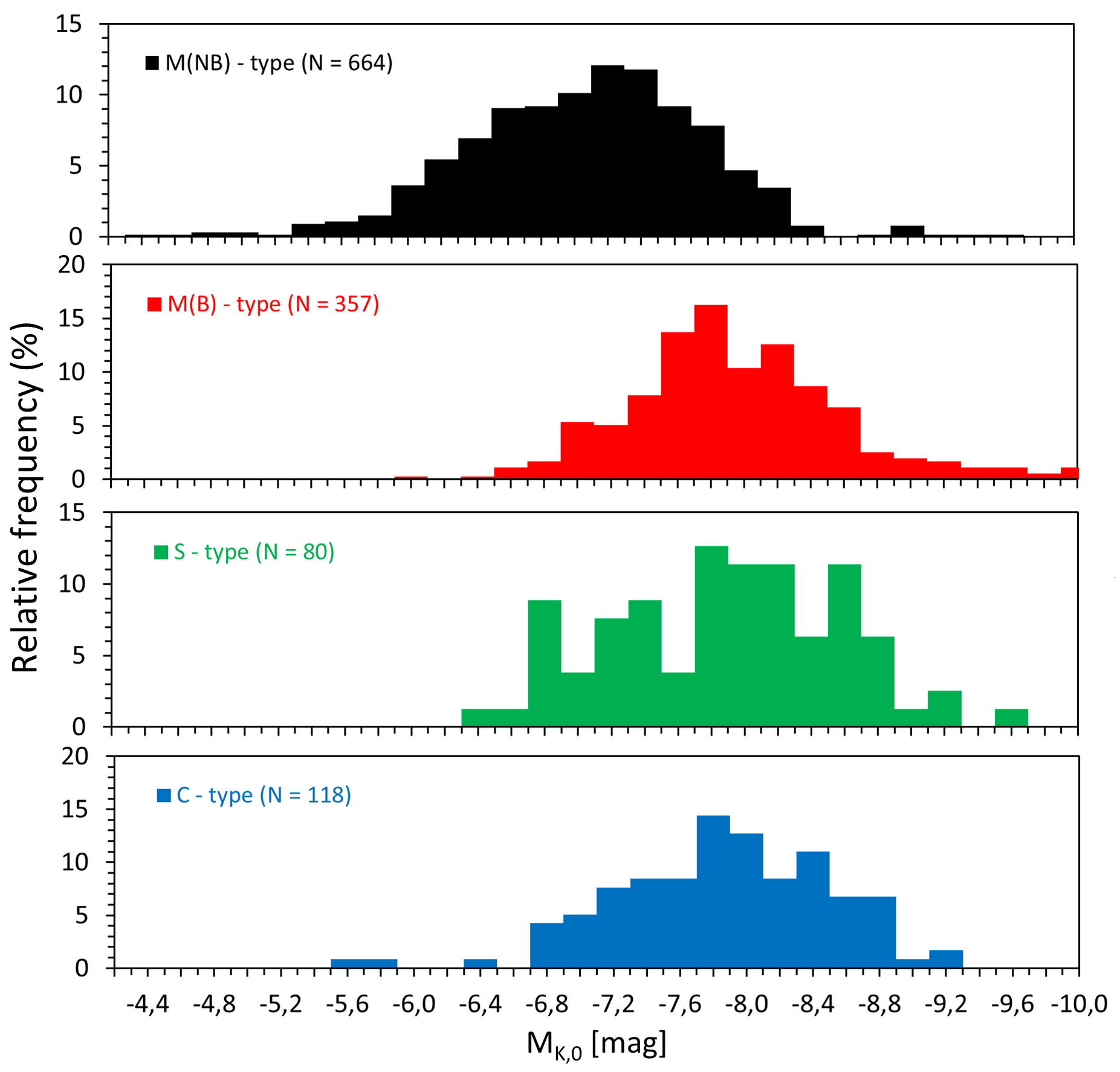}
\centering
\caption{LFs for the Miras of known spectral type in the sample, with bins of 0\fm2. See the legend for the different Mira groups and their number of stars.}
\label{Fig:/Mk Histogram SpT}
\end{figure}

\begin{figure}[t]
\includegraphics[width=\linewidth]{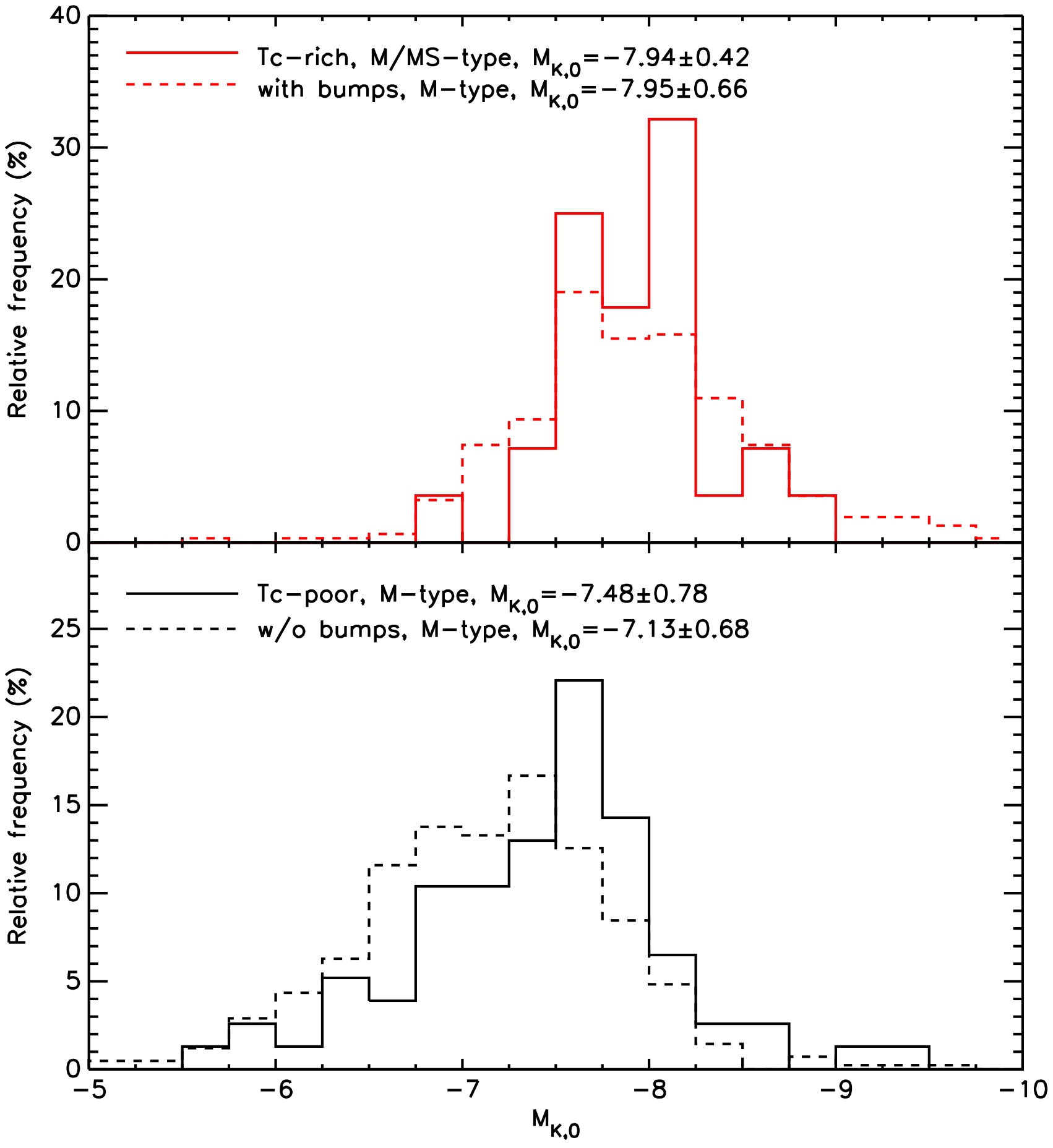}
\centering
\caption{\textit{Upper panel:} $M_{K,0}$ LFs of Tc-rich, M/MS-type Miras from \citet[][solid line]{Uttenthaler2019}, and M-type Miras with bumps in their light curves (dashed line). \textit{Lower panel:} $M_{K,0}$ LFs of Tc-poor, M-type Miras (solid line), and M-type Miras without bumps in their light curves (dashed line). The mean $M_{K,0}$ magnitude and the standard deviation of each group are reported in the legend.}
\label{Fig:MK0_histograms}
\end{figure}

A close relationship was found in Paper~I between the presence of light-curve asymmetries and a possible post-3DUP state in M-type Miras. Here, we made an additional test with our sample and the one used in \citet{Uttenthaler2019}. To be consistent with their sample, only stars that meet the parallax criteria $\varpi>0.4$\,mas and $\varpi/e_{\varpi}>4.0$ were selected for the LFs shown in the $M_{K,0}$ histograms in Fig.~\ref{Fig:MK0_histograms}. In addition, the semi-regular and irregular variables in our sample have not been taken into account, hence the small differences in $\langle M_{K,0} \rangle$ between the Miras with and without asymmetries with respect to those obtained from Fig.~\ref{Fig:/Mk Histogram SpT}. The upper panel of Fig.~\ref{Fig:MK0_histograms} compares the LFs of the Tc-rich M/MS-type Miras of \citet[][28 stars]{Uttenthaler2019} with that of the M(B)-type Miras of our sample (310 stars), and the lower panel compares the LFs of the Tc-poor Miras (77 stars) with that of the M(NB)-type Miras of our sample (414 stars).

The histograms in both panels appear to agree very well. We performed two-sided KS tests to quantify the agreement. The probability that the Tc-rich Miras and those with bumps are drawn from the same parent distribution has a high value of 0.60. The probability that the Tc-poor Miras and those without bumps are drawn from the same distribution, on the other hand, is significantly lower at only $6\times10^{-5}$. Indeed, the distribution of the Tc-poor Miras peaks at a considerably brighter $M_{K,0}$ magnitude than those of the Miras without light-curve bumps. However, it is possible, though difficult to check quantitatively, that the observational works from which the Tc classifications were collected had a selection bias against fainter Miras that are harder to observe in the blue spectral range. We note that Miras with periods $<200$\,d will have masses $<1M_{\sun}$ \citep{Feast2009} and are thus expected to never undergo 3DUP. In any case, a crosscheck between the distributions of the Tc-poor Miras and those with light-curve bumps, as well as between the Tc-rich Miras and those without bumps, also yields low probabilities for an origin from the same distribution: $1.0\times10^{-6}$ and $2.6\times10^{-9}$, respectively.

\subsection{P-L diagrams}\label{subsect: P-L diagrms}

\begin{figure}[t]
\includegraphics[width=\linewidth]{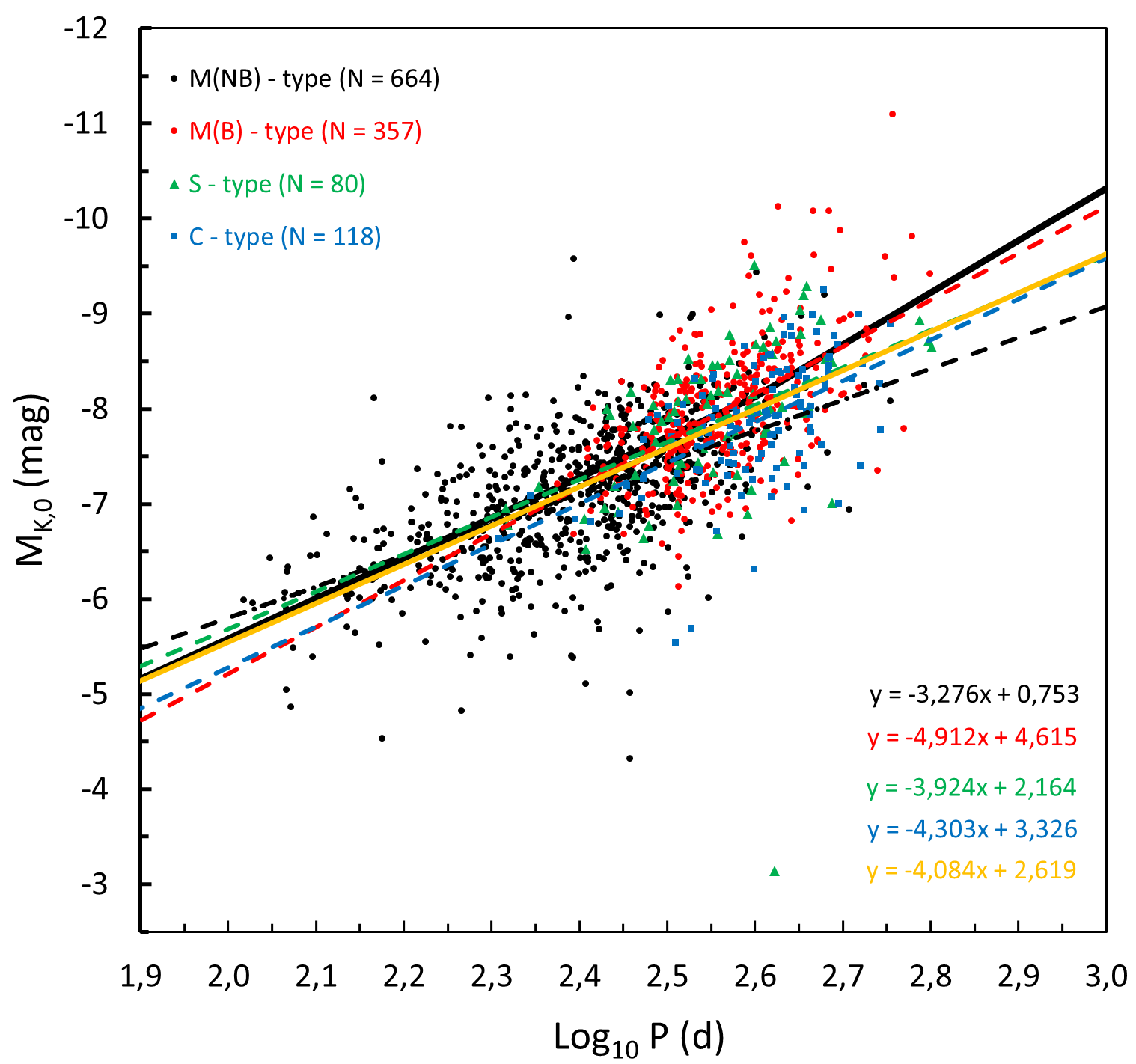}
\centering
\caption{P-L diagram for Miras of known spectral type in the sample. See the legend for the different spectral types, the number of stars in each Group, and the linear fit to each of them. The thick black line indicates the relation obtained by \citet{Sanders2023b} for Galactic Miras. The luminous M(B)-type Miras at $\log P\sim2.6$ could be hot bottom burning candidates.}
\label{Fig:Mk - period SPT}
\end{figure}

It is known that Miras, much like most other LPVs, follow a period-luminosity relation \citep[PLR;][]{Feast1989,Hughes1990}. The sequences depend on the pulsation mode \citep{Whitelock2008,Riebel2010,Wood2015}, and Miras pulsate in the fundamental mode \citep{Trabucchi2017}. We can choose to write for the absolute $K$ magnitude, $M_{K,0}$ \citep{Whitelock2008}: 
\begin{equation}\label{Eq: P-L relation}
M_{K,0} = \rho [log P - 2.38] + \delta \hspace{0.3cm}{\rm [mag]}
,\end{equation}

where $P$ is the Mira pulsation period and $\delta$ the zero point of the PLR with a slope, $\rho$. Figure~\ref{Fig:Mk - period SPT} shows the PL diagram in the $K_{\rm S}$ band for the Mira variables with a known spectral type and Table~\ref{Tab:P-L} gives the results of the linear fits performed on each of the groups. \citet{Whitelock2008} found a slope of $\rho=-3.51$ and a zero point of $\delta=-7.15$ for the O-rich Miras. Due to the period distribution of the Whitelock et al.\ sample ($\sim75\%$ of the Miras had $P\lesssim250$\,d), it is appropriate to compare it with our M(NB) sample (see also Fig.~\ref{Fig:Histogram M - type}). We then see that the slope ($\rho=-3.28$) and the zero point ($\delta=-7.04$) of the M(NB) sample agree well with those found by \citet{Whitelock2008} for their O-rich sample. Details for Miras with an unknown spectral type are given in Appendix~\ref{Appendix A}.

\begin{table}
\caption{Linear P-L fits of the form given by Eq.~\ref{Eq: P-L relation} for the different groups of Miras with a known spectral type in our sample.}
\label{Tab:P-L}
\centering
\begin{tabular}{lccc}
\hline\hline
Group & slope ($\rho$) & Intercept & Zero point ($\delta$)\\
\hline
All stars  & $-4.05\pm0.12$ & $2.54\pm0.30$ & $-7.10$ \\
S-group    & $-3.31\pm0.17$ & $0.83\pm0.41$ & $-7.04$ \\ 
A-group    & $-4.33\pm0.33$ & $3.18\pm0.84$ & $-7.13$ \\ 
M(NB)-type & $-3.28\pm0.18$ & $0.75\pm0.44$ & $-7.04$ \\
M(B)-type  & $-4.91\pm0.39$ & $4.61\pm1.00$ & $-7.08$ \\
S-type     & $-3.92\pm0.94$ & $2.16\pm2.40$ & $-7.17$ \\
C-type     & $-4.30\pm0.64$ & $3.33\pm1.66$ & $-6.92$ \\
W08        & $-3.51\pm0.30$ & $1.20\pm0.06$ & $-7.15$ \\
\hline
\end{tabular}
\tablefoot{Reference: W08 = \citet{Whitelock2008}}
\end{table}

Recently, \citet{Sanders2023b} compared the PLRs of O-rich Miras in the 2MASS $J$, $H$, and $K_{\rm S}$ bands with those measured for the LMC, SMC, the Sagittarius dwarf spheroidal galaxy, globular cluster members, and the Milky Way subset with VLBI parallaxes. Following a linear relation similar to the one given in Eq.~\ref{Eq: P-L relation}, a very good agreement with previous relations \citep[e.g.][]{Whitelock2008,Ita2011,Bhardwaj2019} at short periods ($\log P<2.6$, $P<400$\,d) was found, probably because they also use a linear relation in this regime. From this break period suggested by \citet{Ita2011}, the distribution of magnitudes tends to be steeper than a single linear relation, attributed to the additional luminosity arising from the onset of HBB in stars with $P>400$\,d \citep{Whitelock2003}. We include the relation obtained by \citet{Sanders2023b} for Galactic Miras as a thick black line in Fig.~\ref{Fig:Mk - period SPT}. The PLRs obtained in our sample and displayed in Table~2 show that the slope of the M(B) Miras is significantly steeper than that of the M(NB) Miras. This is consistent with the idea that the PLR of the O-rich Miras is steeper at longer periods, but it is noteworthy that the PLR of the M(B) Miras is even steeper than that of the S- and C-type Miras.

\subsection{Z-scale height and stellar ages}\label{subsect: galactic distribution}

\begin{figure}[t]
\includegraphics[width=\linewidth]{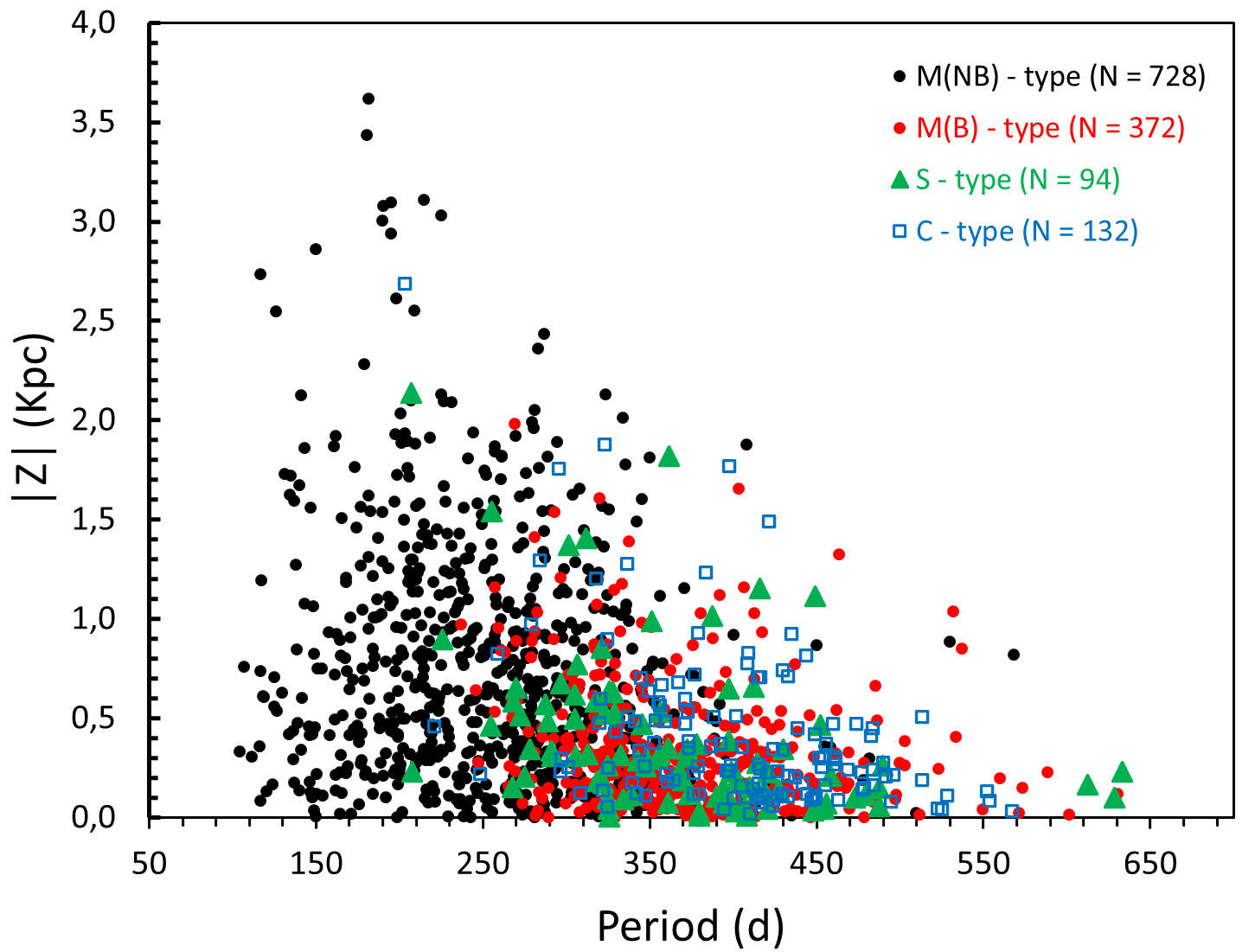}
\centering
\caption{Distance from the Galactic midplane, $|Z|$, as a function of the pulsation period, $P$.}
\label{Fig:Z Midplane}
\end{figure}

The mean distance of a stellar population from the Galactic midplane is a measure of its age, and thus initial stellar mass \citep[e.g.][]{Dove1993}. To analyse possible differences in the $|Z|$ distribution between the O-rich Mira groups in our sample, we constructed a plot of the distance to the Galactic midplane, $|Z|$, versus the pulsation period for the Miras with a known spectral type in Fig.~\ref{Fig:Z Midplane}. The Galactic coordinates have been adopted from the {\it Gaia} DR3 sky coordinates and the distances from \citet{Bailer-Jones2021a}. Most importantly, we see that the mean $|Z|$ and the upper envelope of the $|Z|$ distribution decrease with increasing period. This is a reflection of the period-age relation of Miras \citep{Trabucchi2022}: more massive Miras are younger, and thus have had less time to scatter away from their birth orbit in molecular clouds near the Galactic midplane. Furthermore, we can observe that the M(NB)-type Miras distribute differently in $|Z|$ than the M(B)-, S-, and C-type Miras. The latter hardly reach values of $|Z|\gtrsim1$\,kpc, in contrast to the M(NB)-type Miras that reach values up to $|Z|\approx3.5$\,kpc. The M(B), S, and C Miras, on the other hand, all distribute similarly.

\begin{figure}[t]
\includegraphics[width=\linewidth]{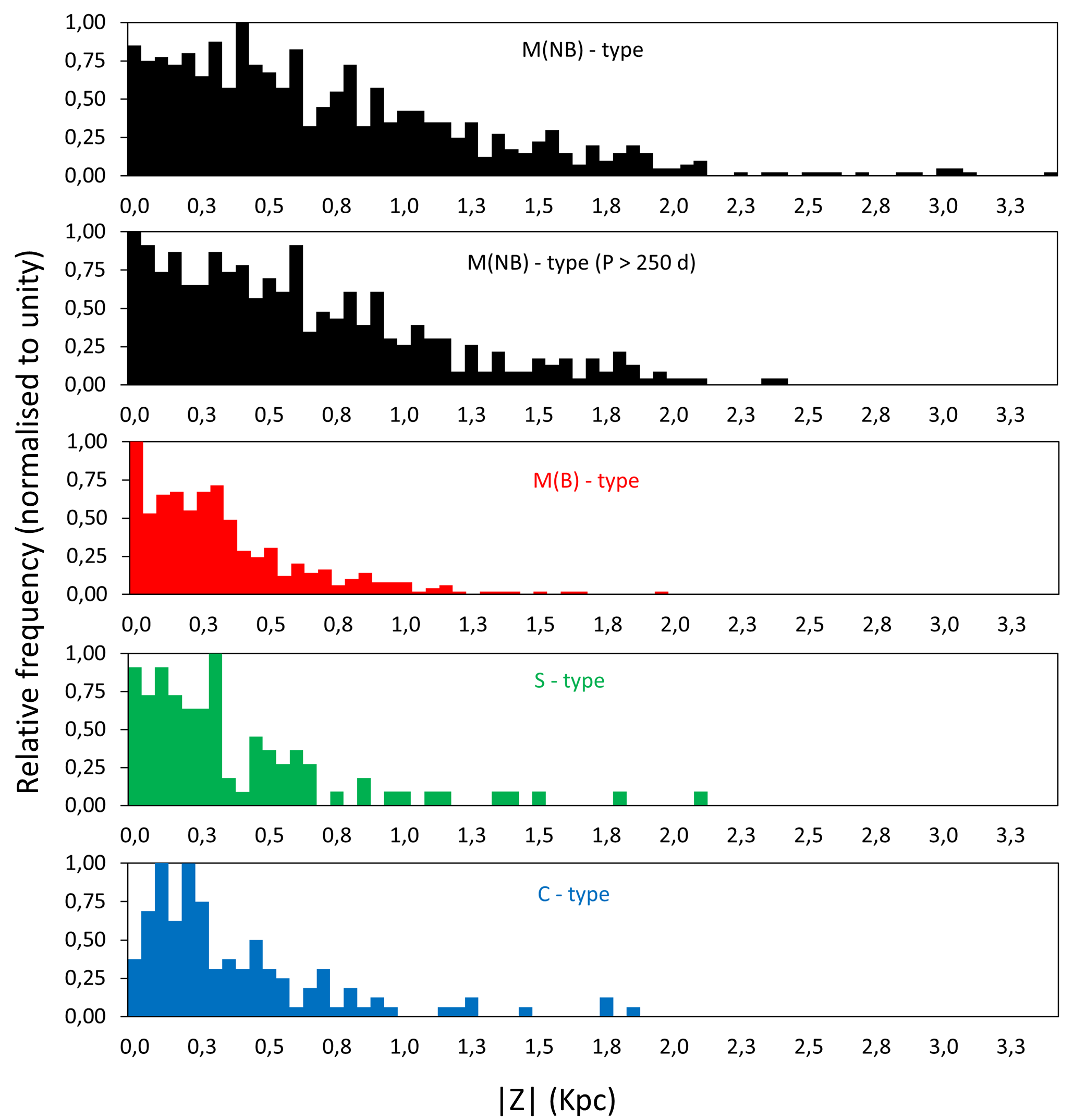}
\centering
\caption{Histogram of the relative frequency of stars as a function of $|Z|$ for the different samples of Miras (see legend).}
\label{Fig:Z Midplane 2}
\end{figure}

We further analyse this point in the histograms in Fig.~\ref{Fig:Z Midplane 2}, where the relative frequency, normalised to a maximum of 1, is plotted as a function $|Z|$. The distribution of the M(B)-type Miras follows a similar pattern to that of the S- and C-type Miras, in contrast to the M(NB)-type Miras. This is reflected in their average distance to the Galactic midplane $\langle|Z|\rangle$, which we report in the second column of Table~\ref{Tab:Galactic distribution}. The M(NB)-type Miras have a value of $\langle|Z|\rangle\approx0.75$\,kpc, whereas the other groups are in the range $0.35\lesssim\langle|Z|\rangle\lesssim0.4$\,kpc. Columns~3 and 4 of the table list the 90th and 95th percentiles of $|Z|$, which follow the same pattern.

The distribution of stars with height above the Galactic midplane can be described by an exponential of the form $N = N_{0} e^{-|Z|/z_{0}}$ \citep{Claussen1987}, where $N$ and $N_{0}$ are related to the spatial density of stars in kpc$^{-3}$, and $1/z_{0}$ is the distance scale. The value of $z_{0}$ can be used to estimate the typical mass of (AGB) star progenitors using tabulations for the scale heights of main-sequence stars as a function of spectral class \citep[e.g.][]{Miller1979}. We use this approach for our sets of Mira-type variables to make an approximate fit to the $|Z|$ distribution with an exponential function \citep[e.g.][and references therein]{Abia2022}. A bin size of 50\,pc was chosen. To cover the entire data, we adopted a range for the fit of $|Z|\lesssim3.5$\,kpc for the M(NB)-type Miras, whereas for the other groups, it was limited to $|Z|\lesssim 2$\,kpc. The result of these fits and the goodness of fit are included in columns 5 and 6, respectively, of Table~\ref{Tab:Galactic distribution}. 

For the C-type Miras, $z_{0}\approx206$ pc is obtained, in good agreement with the range of 150-250\,pc estimated by \citet{Claussen1987}, and with $z_{0}\sim220$\,pc obtained by \citet{Abia2022} for N-type carbon stars. This $z_{0}$ range indicates progenitors with typical main-sequence masses between 1.5 and 2.5\,$M_{\sun}$, fully consistent with the typical AGB carbon star mass \citep[e.g.][]{Straniero2006,Karakas2014}. We find distance scales of $z_{0}=209$ and 227\,pc for M(B)- and S stars, respectively, indicating progenitor masses similar to those of C Miras. On the other hand, for M(NB)-type Miras, a value of $z_{0}\approx506$\,pc is indicative of lower progenitor masses.

In addition, in the second panel of Fig.~\ref{Fig:Z Midplane 2} and the second row of Table~\ref{Tab:Galactic distribution}, we also report the results of the $|Z|$ distribution for the M(NB)-type Miras with $P>250$\,d. We set this limit of $P>250$\,d because M(B), S-, and C-type Miras appear in significant numbers only at periods longer than this threshold. It can be seen how the distribution and scale height of these stars agree with the data obtained for all M(NB)-type Miras, indicating progenitors in the same mass range despite having longer periods, and quite distinct from the M(B)-, S-, and C-type Miras. Therefore, the distance from the Galactic midplane suggests a common origin in terms of the progenitor masses for the M(B)-, S- and C-type stars, from which the M(NB)-type Miras can be distinguished.

\begin{table} 
\caption{Galactic distribution of the five Mira groups.}
\label{Tab:Galactic distribution}
\centering
\begin{tabular}{c c c c c c}
\hline\hline
Type & $\langle|Z|\rangle$ & $|Z|$ & $|Z|$ & $1/z_{0}$ & $R$\\
& (kpc) & 90$^{\rm th}$ p. & 95$^{\rm th}$ p. & (kpc$^{-1}$)\\
\hline
M(NB) & 0.75 & 1.57 & 1.88 & 1.98 & 0.86\\
M(NB) P>250 d & 0.65 & 1.38 & 1.73 & 2.20 & 0.77\\
M(B) & 0.35 & 0.79 & 0.96 & 4.41 & 0.90\\
S & 0.40 & 0.85 & 1.11 & 4.78 & 0.84\\
C & 0.41 & 0.73 & 1.20 & 4.85 & 0.74\\
\hline
\end{tabular}
\tablefoot{Column~1: Mira group designation; column~2: average distance to the Galactic midplane; columns~3 and 4: 90$^{\rm th}$ and 95$^{\rm th}$ percentiles of $|Z|$; column~5: distance scale; column~6: goodness of fit ($R$) to the histograms in Fig.~\ref{Fig:Z Midplane 2}.}
\end{table}

\section{Dust mass-loss rate} \label{sec:dust-MLR}

Near-to-mid infrared colours involving 2MASS and WISE photometry, for example the $K-[22]$ colour, have been proven to trace the (dust) mass-loss rate of evolved giant stars \citep[e.g.][]{McDonald2016,McDonald2018,Uttenthaler2019}. Here, we use the $K-[22]$ and $[3.4]-[22]$ colours as a function of the stars' pulsation period to study their mass-loss properties.

\subsection{$K-[22]$ versus period diagram} \label{Subsect: K22 vs Period}

\begin{figure}[t]
\includegraphics[width=\linewidth]{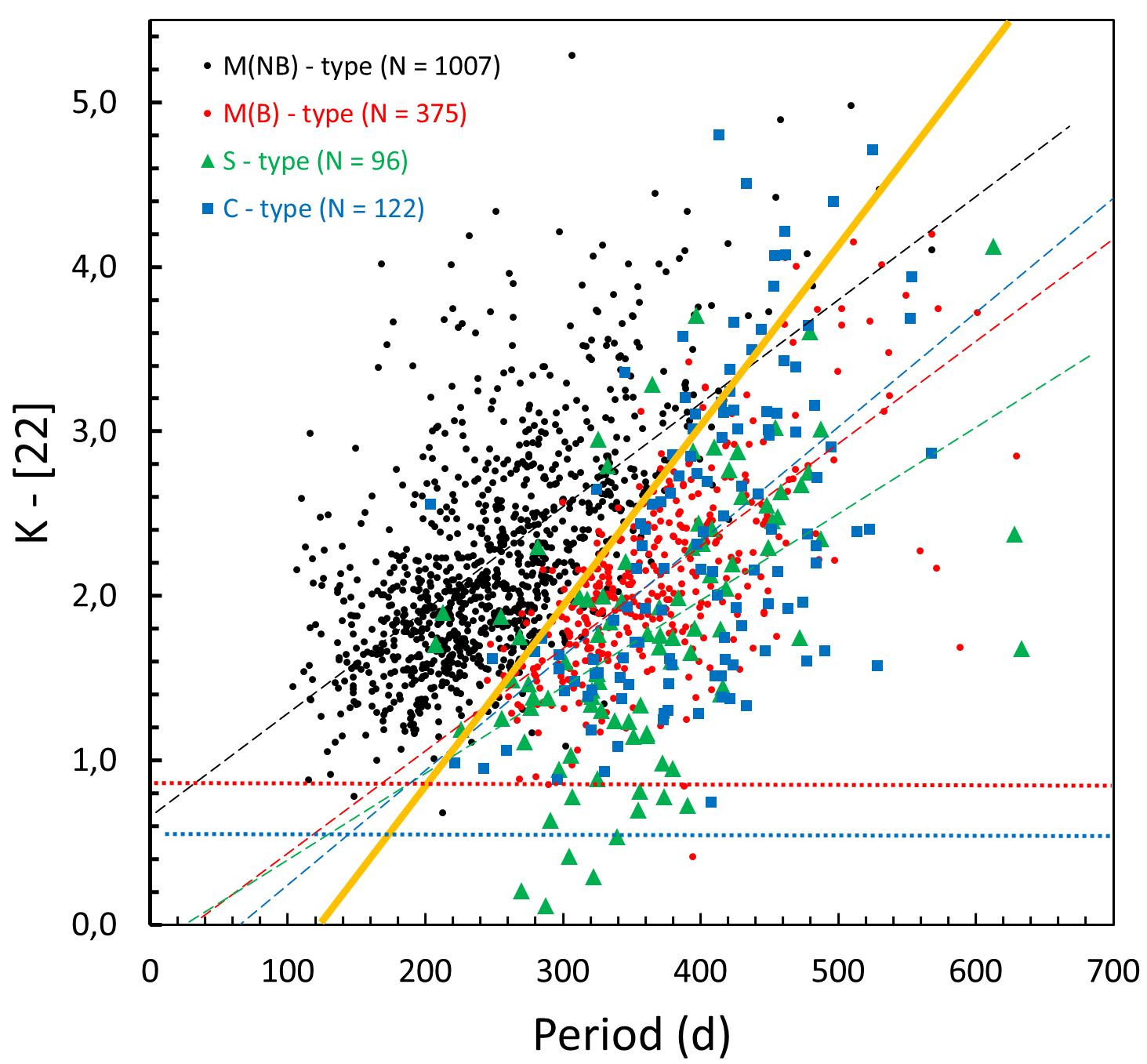}
\centering
\caption{$K-[22]$ vs $P$ diagram of the 1600 stars unsaturated in the $K$ and $[22]$ bands and with a known spectral type; see the legend for identification of the spectral types and the number of Miras in each group. The solid orange line is the relation given by Eq.~\ref{Eq: BSL K-[22]} that best separates Group~S and A Miras. The dashed lines are the linear least-square fits to the four groups of Miras listed in Table~\ref{Tab:Linear fit K-[22]} (black: M(NB), red: M(B), green: S, blue: C Miras). The dotted blue and red lines indicate the limits of $K-[22]=0.55$ and 0.85.}
\label{Fig:K-[22] SpT}
\end{figure}

In our sample, we found a total of 1600 Miras with known spectral types and no saturation in the $[22]$ or $K$ bands (1382 M-type, 96 S-type, and 122 C-type). Details for Miras with an unknown spectral type are included in Appendix~\ref{Appendix A}. The distribution of these stars in the $K-[22]$ versus pulsation period plane is displayed in Fig.~\ref{Fig:K-[22] SpT}. As already demonstrated in Paper~I, we see that the M(NB) Miras accumulate on the upper left, whereas the M(B)-, S- and C-type Miras are on the lower right of the sequence. We implemented an amoeba routine to find the linear relation that best separates the Miras of Groups~A and S by maximising the accuracy of the separation. This yields the relation
\begin{equation}\label{Eq: BSL K-[22]}
K - [22] = 0.011 \times P - 1.380  \hspace{0.3cm}{\rm [mag]},
\end{equation}
which, surprisingly, is the exact same linear relation found by \citet{Uttenthaler2019} that best separates Tc-poor and Tc-rich Miras in that diagram. The accuracy of the separation, i.e. the probability that a star is located on the 'correct' side of the line, is even slightly higher in this case (0.89) than was found for the separation of Tc-poor and Tc-rich Miras (0.87). The relation is included in Fig.~\ref{Fig:K-[22] SpT} as a solid line. The fact that the stars with asymmetries concentrate in the same region of the diagram as those with Tc, and the ones without asymmetries in the same region as those without Tc, again suggests a possible connection between the two properties.

The dotted blue line in Fig.~\ref{Fig:K-[22] SpT} indicates the $K-[22]=0.55$ limit set by \citet{McDonald2018}, above which cool giant stars are detectable in relatively deep CO millimetre observations. Stars above this limit could have $\dot{M}\gtrsim8\times10^{-8}M_{\sun}{\rm yr}^{-1}$. The dotted red line, on the other hand, is the $K-[22]=0.85$ limit established in \citet{McDonald2016}, above which significant dust production and mass loss set in, related to the pulsation period $P\gtrsim60$\,d. We note that there is a significant fraction of S-type Miras ($11/95=12$\%) below this second limit and thus seem to be relatively dust-free, despite their long pulsation periods of $\gtrsim300$\,d.

For the M-type Miras, the distribution of the data reveals that $375/1382\approx27.1\%$ of them are of type M(B), in good agreement with the $\sim31\%$ obtained in Paper~I and agrees excellently with the fraction of Tc-rich stars found there, namely 26.7\%. Its distribution around the line given by Eq.~\ref{Eq: BSL K-[22]} yields that 96.1\% of the M(NB) Miras lie above this line, while 91.7\% of the M(B) Miras are concentrated below this line. Moreover, most of those stars that are not located in the 'correct' zone are very close to the separation line. Regarding S- and C-type Miras, the same trend is observed, with quite significant, although smaller, fraction of stars located in the correct region of the diagram, with 88.4\% and 82.8\%, respectively. In Fig.~\ref{Fig:K-[22] vs Period} in the Appendix we can check in detail the distribution of the different Mira groups around the dividing line given by Eq.~\ref{Eq: BSL K-[22]}. It is interesting to see how the Miras with unknown spectral types also follow this same sequence depending on whether or not they have asymmetries in their light curves. 

We also see in Fig.~\ref{Fig:K-[22] SpT} how the $K-[22]$ value and, thus, the dust mass-loss rates, increase with increasing pulsation periods. Nevertheless, the sequences formed by the M(NB)- and M(B)-Miras are very different, the latter starting at $P\gtrsim250$ days but at lower $K-[22]$ values than for the M(NB) Miras. The linear fits to the groups of Miras are plotted in Fig.~\ref{Fig:K-[22] SpT} as dashed lines with the same colour as the respective symbols, and their coefficients are collected in Table~\ref{Tab:Linear fit K-[22]}. The fits are valid for an approximate period range of $100\lesssim P/{\rm d}\lesssim400$ for M(NB), and $240\lesssim P/{\rm d}\lesssim640$ for M(B), S, and C types. The colour slopes have very similar values, but the intercept of the M(NB) Miras differs by $\gtrsim0\fm7$ from the other groups. \citet{Uttenthaler2019} determined a linear relationship between $K-[22]$ colour and $\log(\dot{M}_{g}/M_{\sun})$ to estimate the gas mass loss rate of AGB stars, applicable approximately in the interval $1.5\lesssim K-[22]\lesssim7.5$, which has the form: $\log(\dot{M}_{g}/M_{\sun})=0.392\times(K-[22])-7.431$. At $P=300$\,d, the M(B) and M(NB) Miras differ by $\sim0.85$\,mag in $K-[22]$. Applying this to the relation between the $K-[22]$ index and the gas mass-loss rate, we find that the M(B) Miras have a gas mass-loss rate lower than that of the M(NB) Miras by more than a factor of 2.

\begin{table} 
\caption{Parameters of the linear fit $K-[22]=a\times P+b$ in the $K-[22]$ diagram for the different groups of Miras in the sample.}
\label{Tab:Linear fit K-[22]}
\centering
\begin{tabular}{cccc}
\hline\hline
Type & $a$ & $b$ & r\\
\hline\
M(NB) & $0.00628\pm0.00025$ & $+0.654\pm0.068$ & 0.61\\
M(B)  & $0.00623\pm0.00033$ & $-0.190\pm0.122$ & 0.71\\
S     & $0.00526\pm0.00085$ & $-0.132\pm0.318$ & 0.54\\
C     & $0.00696\pm0.00101$ & $-0.453\pm0.414$ & 0.53\\
\hline
\end{tabular}
\end{table}

\subsection{$[3.4]-[22]$ diagram}\label{subsect: W1-W4 vs Perios}

Instead of the $K$-band magnitude, we can also use the [3.4] WISE band. It has the advantage over the $K$ band in that it is measured simultaneously with the [22] band and that Miras have smaller amplitudes of variability at that wavelength, both of which should reduce the scatter in the colour index. For stars that are not excessively dusty, the [3.4] band is still dominated by photospheric emission, and we can adopt the $[3.4]-[22]$ colour as a similar indicator of the dust mass-loss rate. In our sample, we found a total of 1065 Miras unsaturated in the [3.4] and [22] bands with known spectral types (953 M-type, 49 S-type, and 63 C-type; see Appendix~\ref{Appendix A} for details of the study on Miras with an unknown spectral type.) Again, we optimised a linear relation that best separates Group~A and S Miras using the same process as in section~\ref{Subsect: K22 vs Period}, from which we find the relation:
\begin{equation}\label{Eq: BSL W1-W4}
[3.4] - [22] = 0.0086 \times P - 1.0093  \hspace{0.3cm}{\rm [mag]}.
\end{equation}

\begin{figure}[t]
\includegraphics[width=\linewidth]{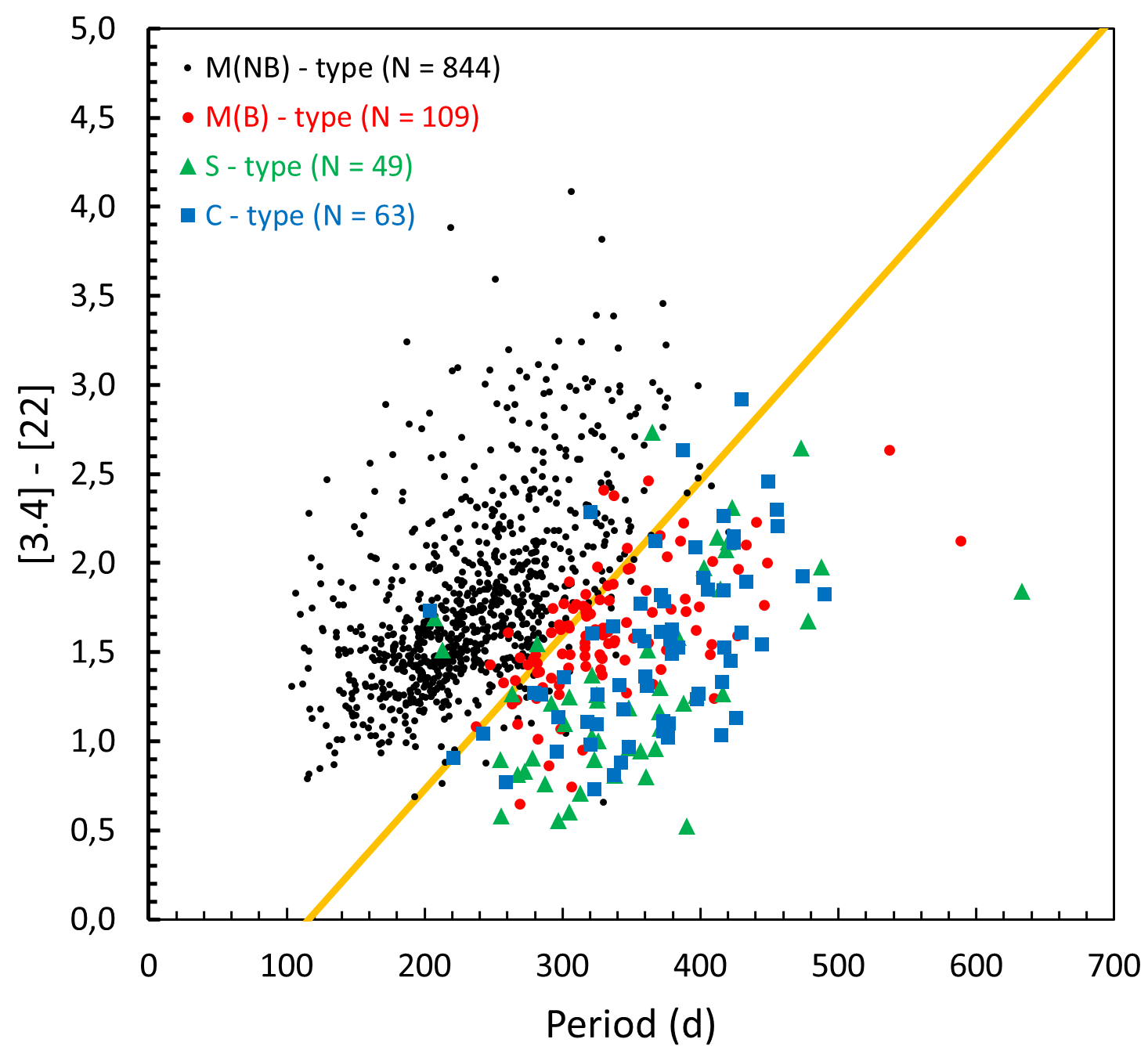}
\centering
\caption{$[3.4]-[22]$ vs $P$ diagram of the 1065 stars in our sample with a known spectral type and no saturation in both the [3.4] and [22] bands; see legend for identification of the different spectral types and the number of Miras in each group. The solid orange line is the relation that best separates Group~S and A Miras given in Eq.~\ref{Eq: BSL W1-W4}.}
\label{Fig:W1-W4 SpT}
\end{figure}

The accuracy of this linear relation is 0.873, which means that the probability that a star is on the 'correct' side of the relation in the $[3.4]-[22]$ versus $P$ diagram is 87.3\%. Figure~\ref{Fig:W1-W4 SpT} shows the $[3.4]-[22]$ versus period diagram for the sample Miras with a known spectral type, including the relation given by Eq.~\ref{Eq: BSL W1-W4} with a solid orange line. Many similarities with the $K-[22]$ versus $P$ diagram (Fig.~\ref{Fig:K-[22] SpT}) can be seen. An even larger percentage of the M(NB) Miras lie above this dividing line than the corresponding one in the $K-[22]$ versus $P$ diagram, namely 97.3\%. Also, many of the S- and C-type Miras are below the dividing line, 91.8\% and 92.1\%, respectively. The same trend is observed for the M(B) Miras as for the S- and C-types, although with a lower fraction of stars located below the dividing line, namely 76.1\%. We present in more detail in Fig.~\ref{Fig:W1-W4 vs period} in the appendix the distribution of the different Mira groups around the dividing line given by Eq.~\ref{Eq: BSL W1-W4}.

We add a note of caution here because saturation introduces some bias here. As can be seen in Fig.~\ref{Fig:W1-W4 SpT}, there is a paucity of stars with $P\gtrsim450$\,d. As long-period stars have higher intrinsic luminosity, there is also a higher probability that they saturate in one of the involved bands, [3.4] or [22]. A fraction of 39.4\% of the stars are saturated in one of the bands. This loss of sample stars is not evenly distributed over the different spectral types. Significantly more M(B)-, S-, and C-type stars are saturated in one or both two bands (73.1\%, 55.4\%, and 54.3\%, respectively) than among the M(NB) Miras (23.8\%). This saturation bias need not necessarily impact the relation found in Eq.~\ref{Eq: BSL W1-W4}, but it removes a non-negligible fraction of stars from Fig.~\ref{Fig:W1-W4 SpT}.

\section{{\it Gaia}-2MASS photometry}\label{sec: Gaia-2MASS photometry}

\subsection{The {\it Gaia}-2MASS diagram}

The intrinsic colour dispersion of red giants is relatively small in the IR, except for strong circumstellar reddening caused by dust ejected by some stars. A reddening-free magnitude can be obtained from 2MASS photometry \citep{Skrutskie2006} using the Wesenheit function $W_{K}$ in the NIR \citep{Soszynski2005}. It takes the form $W_{K_{\rm S},J-K_{\rm S}} = K_{\rm S} - 0.686(J-K_{\rm S})$, since interstellar reddening impacts both the $K_{\rm S}$ magnitude and the $J-K_{\rm S}$ colour. In contrast, in the visual range, red giants show a large intrinsic colour dispersion, specifically in $G_{BP} - G_{RP}$. The molecular absorption in these bands is very sensitive to the stellar temperature and the C/O ratio. Nevertheless, a combination of the $G_{BP}$ and $G_{RP}$ magnitudes can be obtained using the Wesenheit function $W_{RP}$, which eliminates the interstellar extinction (and much of the circumstellar extinction, as long as the interstellar and circumstellar extinction laws are similar). It takes the form $W_{RP,RP-BP} = G_{RP} - 1.3(G_{BP}-G_{RP})$ \citep{Lebzelter2018}.

The $W_{RP}$ function combines information on temperature, chemistry, reddening, and brightness of LPVs, while the $W_{K}$ function is much less sensitive to both photospheric temperature and chemistry. Combining both functions, we can construct the $W_{RP}-W_{K_{_S}}$ versus $M_{K,0}$ diagram, which is a powerful tool to analyse and identify sub-classes of LPVs, both as a function of the chemical type and initial mass, initially designed by \citet{Lebzelter2018} in a study of LPVs in the LMC. The {\it Gaia}-2MASS diagram for the Miras with a known spectral type in our sample is shown in Fig.~\ref{Fig:WRP-WK vs MK SpT} (see Appendix~\ref{Appendix A} for the Miras with an unknown spectral type). The various branches (groups) of stars as defined by \citet{Lebzelter2018} are delineated in that figure. 

\begin{figure}[!t]
\includegraphics[width=\linewidth]{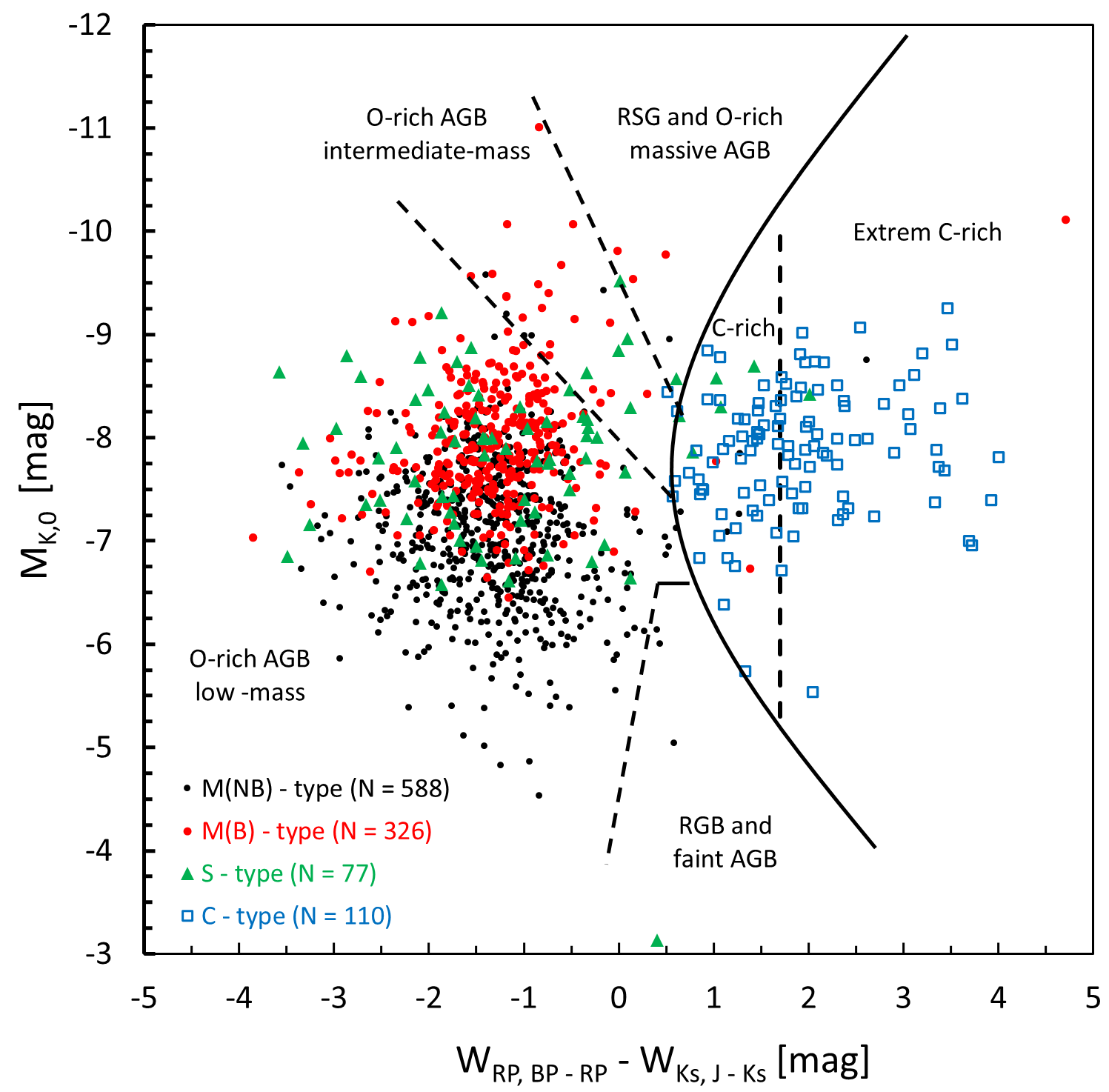}
\centering
\caption{$W_{RP,RP-BP}$ vs\ $M_{K,0}$ for stars with a known spectral type. See legends to identify the Mira type and the number of stars in each group.}
\label{Fig:WRP-WK vs MK SpT}
\end{figure}

\subsection{New DBSL parameter} \label{sec: DBSL function}

We now propose a new parameter that we call the distance to the best separation line (DBSL), defined as the distance between the observed colour index of a Mira from the line that best separates the Group~A and S Miras in that colour index versus period $P$ (Eq.~\ref{Eq: BSL K-[22]} and Eq.~\ref{Eq: BSL W1-W4}), given by the expressions:
\begin{equation}\label{Eq: DBSL (k-[22])}
\delta_{K-[22]} = (K - [22]) - (0.011 \times P - 1.380) \hspace{0.3cm}{\rm [mag]}
,\end{equation}
\begin{equation}\label{Eq: DBSL (W1-W4)}
\delta_{[3.4]-[22]} = ([3.4]-[22]) - (0.0086 \times P - 1.0093) \hspace{0.3cm}{\rm [mag]}
.\end{equation}
 
Group~S Miras would generally have positive values of the DBSL parameter, while most Group~A Miras would have negative values (see Figs.~\ref{Fig:K-[22] SpT} and~\ref{Fig:W1-W4 SpT}). 

\begin{figure*}[t]
\includegraphics[width=\linewidth]{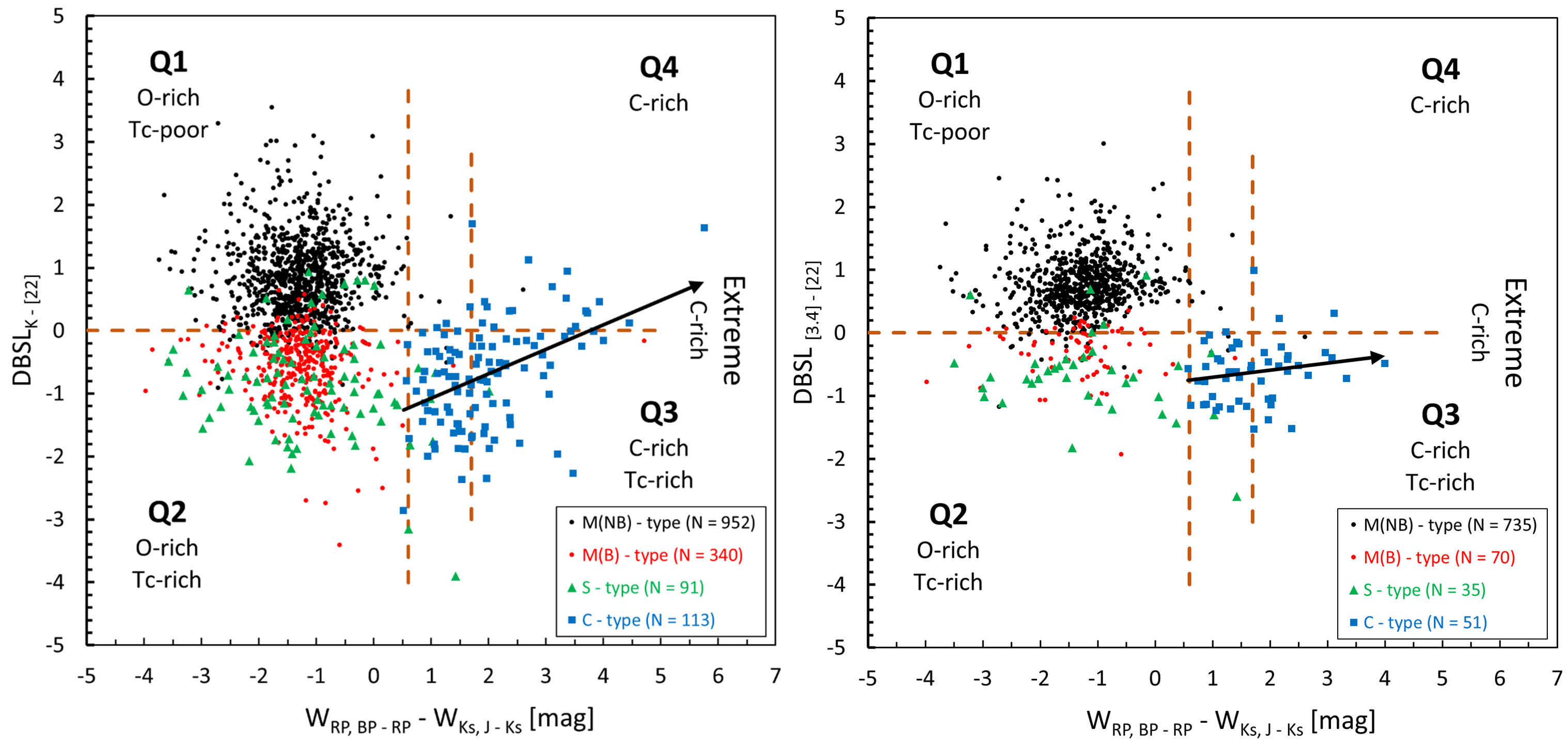}
\centering
\caption{\textit{Left panel:} $\delta_{K-[22]}$ vs\ $W_{RP} - W_{K}$ diagram. \textit{Right panel:} $\delta_{[3.4]-[22]}$ vs\ $W_{RP} - W_{K}$ diagram. See legends to identify the Mira type and the number of stars in each group.}
\label{Fig:DELTA SPT}
\end{figure*}

We constructed new diagrams, $\delta_{K-[22]}$ versus $W_{RP} - W_{K}$ and $\delta_{[3.4]-[22]}$ versus $W_{RP} - W_{K}$, which combine the advantages of the colour index $K-[22]$ and $[3.4]-[22]$ used in the DBSL parameters, and the Wesenheit index $W_{RP} - W_{K}$. These diagrams are shown in Fig.~\ref{Fig:DELTA SPT}. We divided these diagrams by framing four quadrants, separated by the horizontal line for $DBSL = 0$ and by the vertical line $W_{RP} - W_{K} = 0.6$, to the right of which the C-type Miras are preferentially located. We also included the line at $W_{RP} - W_{K} =1.7$ that separates the extreme C-rich stars. The stars in those quadrants have the characteristics described in the following.

Q1, $DBSL>0$, $W_{RP}-W_{K} \lesssim 0.6$: This area is occupied mainly by M-type Miras with symmetric light curves (Group~S) and relatively high dust mass-loss rates. A fraction of 95.8\% of the stars in Q1 are M(NB)-type. They are generally fainter than the other groups ($-8.0\lesssim M_{K,0}\lesssim-5.5$) and are distributed on average at relatively large distances from the Galactic midplane ($\langle|Z|\rangle\approx0.75$\,kpc).

Q2, $DBSL<0$, $W_{RP}-W_{K} \lesssim 0.6$: This quadrant contains mainly stars with light-curve asymmetries (Group~A) that have lower dust mass-loss rates than their Q1-quadrant siblings at comparable pulsation periods, as suggested by their bluer $K-[22]$ colours (see Fig.~\ref{Fig:K-[22] SpT}). A fraction of 88.9\% are M(B)- and S-type Miras. They are brighter on average than the M(NB)-type Miras in Q1 ($-9.5\lesssim M_{K,0}\lesssim-6.8$), and are distributed closer to the Galactic midplane ($0.35\lesssim\langle|Z|\rangle\lesssim0.4$\,kpc). Most of the stars in Q2 are expected to have 3DUP activity.

Q3-Q4, $W_{RP}-W_{K} \gtrsim 0.6$: These two quadrants are occupied mainly by C-type stars (88.7\%). A linear trend is observed that points from Q3 to Q4 as stars with longer periods are located increasingly above the line that best separates the Group~A and S Miras (Fig.~\ref{Fig:K-[22] SpT}). Their higher reddening indicates higher dust mass-loss rates, as has also been demonstrated with comparable near-IR colours by \citet[][their Fig.~15]{Suh2020}. The stars in Q3 and Q4 distribute similarly in luminosity and distance from the Galactic midplane as the M(B)- and S-type stars in Q2.

Thus, the DBSL versus $W_{RP}-W_{K}$ diagram efficiently separates not only O-rich from C-rich stars but also those with light-curve asymmetries from those without. Each of the quadrants marks a different evolutionary stage on the AGB through the sequence $Q1 - Q2 - Q3 - Q4$, corresponding to the evolutionary sequence $M-S-C-({\rm extreme C})$. While the Q1 quadrant is essentially composed of low-mass stars with symmetric light curves and without 3DUP activity, the Q2 quadrant would preferentially be composed of stars with 3DUP activity, S- and M(B)-type Miras with masses in the same range as C-type Miras, $1.4\lesssim M_{i}/M_{\sun}\lesssim3.2$ \citep{Lebzelter2018}. The minimum mass of carbon star formation is a much sought-after question, and new computations by \citet{Rees2024} suggest a limit at $1.5-1.75M_{\sun}$.

\section{New carbon star candidates} \label{sec: New carbon stars}

As we found a few discrepancies in spectral-type classifications between literature sources and the new observational information for some of the stars, we ran a systematic check of the classifications. We present the results of this exercise in this section.

The only reliable way to identify whether a star is O-rich or C-rich (${\rm C}/{\rm O}<1$ or ${\rm C}/{\rm O}>1$, respectively) is through its optical spectrum, identifying absorption bands of oxides (e.g. TiO, VO, LaO, ZrO) or carbon-bearing molecules (e.g. CN, C$_{2}$, CH), respectively. Photometric methods are not as reliable, although they can give good results if previously calibrated on sets of stars with O-rich or C-rich nature known from spectroscopy or if they are based on specifically designed narrow filter systems \citep{Palmer1982}. Moreover, these stars can change from O-rich to C-rich in a relatively short time, so it is not excluded that outliers in their colour indices may be found due to photometry taken at different epochs \citep{Uttenthaler2016b}. 

\subsection{{\it Gaia} Spectra} \label{Subsect:Gaia BP/RP spectra}

Despite its low resolution \citep[$\langle R_{\lambda}\rangle\sim50$, in the range $30-100$ for BP and $70-100$ for RP,][]{DeAngeli2023}, {\it Gaia} DR3 RP spectra enable us to distinguish the chemical types based on the presence of bands of oxides or carbon molecules. \citet{Lebzelter2023} developed a classification based on the pseudo-wavelength peak separation median\_delta\_wl\_rp (hereafter \texttt{mdwlrp}). Specifically, \texttt{mdwlrp} is the distance between the two highest peaks.

The BP/RP spectra of M-type AGB and RSG stars are dominated by absorption due to TiO bands, mainly at 705.4, 758.9, 819.4, 843.2, 885.9, and 920.9\,nm \citep[e.g.][and references therein]{Sanders2023a}, which give rise to characteristic peaks at about 700, 750, 820, 875, and 920\,nm (the latter are often mixed in the BP/RP spectra). Furthermore, in the spectra of the cooler Miras stars, VO features appear at 740, 790, and 800\,nm. They broaden the other oxide bands in their vicinity, sometimes even presenting a double minimum structure as observed around 780\,nm in some Miras, one formed by TiO and one by VO. The BP/RP spectra of S-type stars show two distinct absorption bands by ZrO at 653 and 940\,nm \citep[e.g.][and references therein]{Messineo2023}. The most characteristic is the one at 940\,nm \citep[e.g.][]{Rayner2009, Messineo2021}. The TiO and VO bands are generally weaker than in M-type stars. The two LaO absorption bands 740.6 and 791.4\,nm can appear as an absorption feature centred on 790\,nm. Finally, the BP/RP spectra of C-type Mira stars show three very noticeable peaks at 683, 778, and 900\,nm due to the characteristic CN series \citep[e.g.][]{Kraemer2005, Gonneau2016}, and notable minima at approximately 704.8, 803.3, and 932.8\,nm \citep[][with a model based on the BP/RP spectra of eleven late S-type stars]{Messineo2023}.

The value of \texttt{mdwlrp} is used to define the 'isCstar' parameter \citep{GaiaCollaboration2022}, which is set to 'TRUE' if \texttt{mdwlrp}$>7$, 'FALSE' if \texttt{mdwlrp}$<7$, and 'NULL' when the shape of the spectrum does not allow an automatic classification of these two types. The \texttt{mdwlrp} parameter is reliable for discriminating between O-rich and C-rich stars as long as the signal-to-noise ratio is sufficiently high. \citet{Lebzelter2023} suggest relying on the C-rich classification only if $7<\texttt{mdwlrp}<11$ and $G_{BP}<19$.

We constructed a discriminatory model for most of the absorption bands and peaks discussed above to distinguish between the chemical spectral types. For this model, we selected the ten stars of each of the three spectral types with the highest fluxes, ensuring that the signal-to-noise ratio is optimal and that the spectra are not saturated. For each of the absorption bands and associated peaks, the average wavelength at which they occur has been determined. Table~\ref{Tab:average wavelength} shows these results, with their type (peak or minimum) in column 1, the average wavelength where they occur and their standard deviation in columns 2 and 3, respectively, as well as the molecular band with which they are associated in column 4. Finally, the last column shows 'yes' or 'no' to indicate whether they are used or not in the subsequent discriminatory model between O-rich and C-rich, as we describe below.

\begin{figure*}[!t]
\includegraphics[scale=0.45]{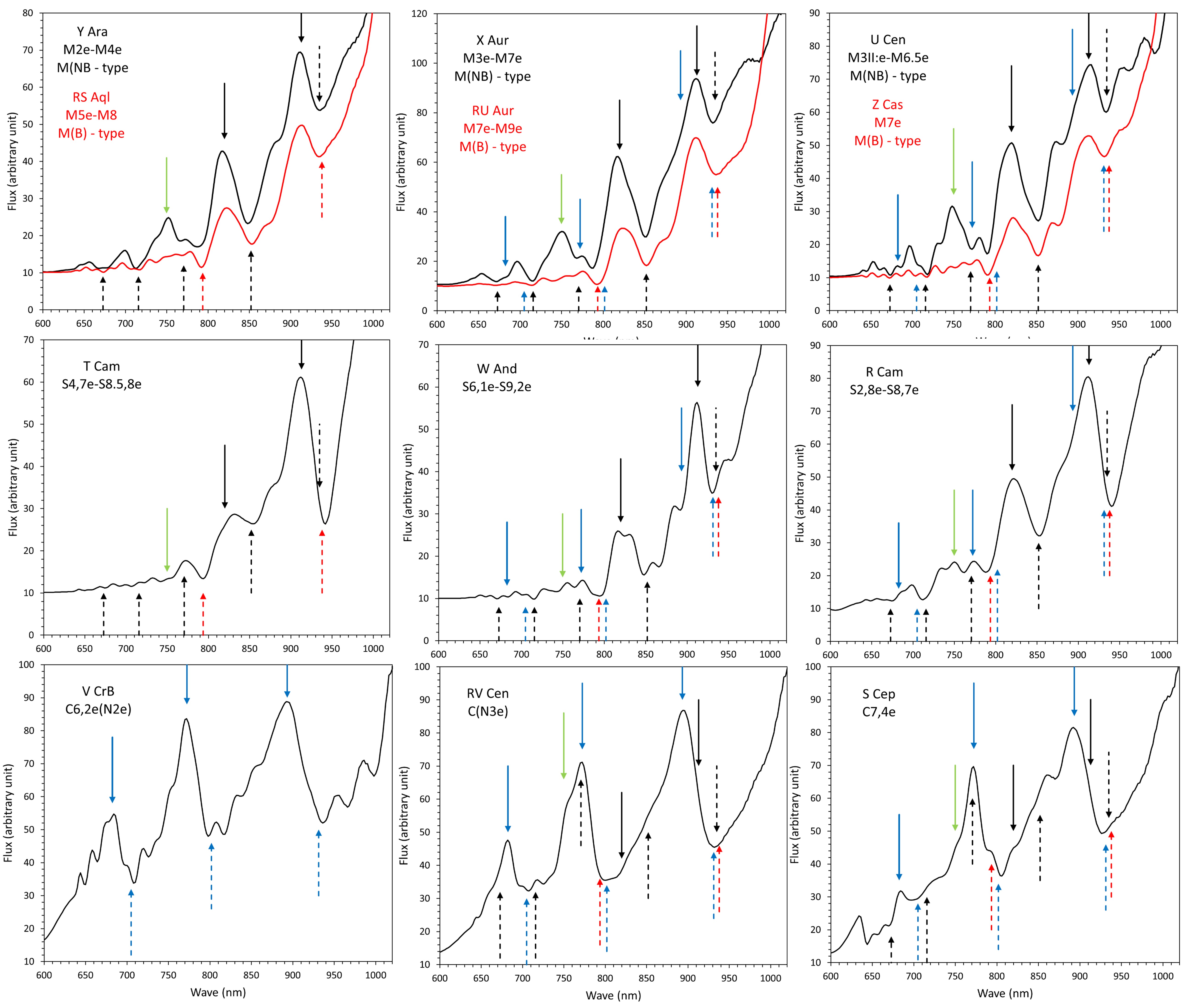}
\centering
\caption{{\it Gaia} BP/RP spectra of three stars per chemical spectral type used to construct the discriminatory model. First row: M(NB)- and M(B)- type; second row: S-type; third row: C-type. Solid and dashed arrows mark peaks and minima, respectively, of characteristic molecular bands: TiO (black); LaO, VO, or ZrO (red), and CN (blue). The solid green arrow shows the peak of the TiO absorption bands at 749.8\,nm.}
\label{Fig:Spectra 0}
\end{figure*}

For S and C spectral types, the BP/RP spectra of three stars among those selected to obtain the model are shown in Fig.~\ref{Fig:Spectra 0}. For M-type stars, three M(NB)-type Miras and three M(B)-type Miras are displayed. The average wavelengths corresponding to the 'peaks' are shown with downward arrows, while those corresponding to the 'minima' of the absorption bands are shown with upward arrows. In Fig.~\ref{Fig:Spectra 0}, the main average wavelengths of the TiO absorption bands characteristic of O-rich Miras are marked, as well as those corresponding to VO, LaO, and ZrO (the latter two being characteristic of S-type Miras), and those corresponding to the CN absorption bands (characteristic of C-type Miras). The peak of the TiO absorption bands at 749.8\,nm differs between the M(NB) and M(B) Miras. Higher resolution spectra would be needed for a satisfactory distinction of M(NB) and M(B) Miras. The M(B) spectra resemble the S stars at this peak, which could mean that 3DUP activity influences this peak. However, we also remark that the M(B) Miras appear to be of a much later subtype than the M(NB) Miras plotted in Fig.~\ref{Fig:Spectra 0}. One also must not forget that the {\it Gaia} BP/RP DR3 spectra are averaged over several epochs and that spectral features may be smeared out for large amplitude pulsators.

In this figure, it can be seen how the three peaks corresponding to the CN absorption bands are prominent in the C-rich Miras, whereas they form part of the pseudo-continuum for the O-rich Miras. Vice versa, it is observed that the two features associated with the TiO and ZrO absorption bands are prominent in the M and S Miras, respectively, whereas they are part of the pseudo-continuum of the C-rich Miras. The peak due to TiO absorption bands above $\sim750$\,nm is not included in the discriminatory model because, although it does not appear in any C-rich Miras, it appears only in some O-rich Miras. As for the minima, the one corresponding to the TiO bands redward of 850\,nm stands out, which for O-rich stars is generally very deep. We used this set of six average wavelengths in our discriminatory model between O-rich and C-rich stars (Table~\ref{Tab:average wavelength}).

\begin{table}
\caption{Different absorption bands and associated peaks derived from our discriminatory model.}
\label{Tab:average wavelength}
\centering
\begin{tabular}[b]{c c c c c}
\hline \hline
Feature & $\langle \lambda \rangle$ (nm) & $\sigma$ (nm) & Bands & Usability\\
\hline
Minimum & 672.8 & 1.4 & TiO & no\\
Minimum & 715.8 & 2.5 & TiO & no\\
Peak & 749.8 & 1.2 & TiO & no\\
Minimum & 770.5 & 1.9 & TiO & no\\
Minimum & 793.5 & 1.7 & VO-LaO & no\\
Peak & 819.8 & 0.7 & TiO & yes\\
Minimum & 852.0 & 2.0 & TiO-VO & yes\\
Peak & 913.0 & 2.2 & TiO & yes\\
Minimum & 934.8 & 2.6 & TiO & no\\
Minimum & 937.8 & 3.4 & ZrO & no\\
Peak & 682.5 & 0.9 & CN & yes\\
Minimum & 704.9 & 2.9 & CN & no\\
Peak & 772.3 & 0.7 & CN & yes\\
Minimum & 802.0 & 3.4 & CN & no\\
Peak & 893.3 & 2.0 & CN & yes\\
Minimum & 931.3 & 3.5 & CN & no\\
\hline\end{tabular}
\end{table}

For the rest of the minima corresponding to the different absorption bands, we found that they will not be effective in discriminating between O-rich and C-rich atmospheres because of the proximity of several absorption bands in a wavelength interval too narrow. However, they will generally be used to corroborate the classification derived from the BP/RP spectra. Furthermore, it is observed that three minima corresponding to bands of TiO, CN, and ZrO are very close to each other in the region $930-940$\,nm, so this region is almost useless for the discriminatory model at the low resolution of the BP/RP spectra. Therefore, they will not be taken into account either to discriminate between O-rich and C-rich or to confirm the possible spectral classification of these spectra.

\subsection{Discriminatory model and individual objects}\label{Discriminative diagnosis}

With the above considerations, our discriminative model will be composed of three criteria that have been shown to conveniently distinguish between O-rich and C-rich stars. Firstly, the main peaks and minima of the molecular bands in the BP/RP spectrum (see Table~\ref{Tab:average wavelength}), which discriminate very effectively. Secondly, the $W_{RP}-W_{K}$ index, which separates between O-rich and C-rich stars with a threshold value of $W_{RP}-W_{K}\approx0.6$, although some C-rich Miras may be very close to the threshold. Third, the 'isCstar' flag ($=1$ for C-rich stars and $=0$ for O-rich stars), derived from the \texttt{mdwlrp} parameter, although there can be false positive and negative classifications \citep[see][for a more detailed discussion]{Lucey2023}. The values of these parameters for the candidate carbon stars are shown in Table~\ref{Tab:C-type candidate Miras}, for both Miras with known and unknown spectral types, and these discriminative criteria yield the following results:

\begin{itemize}
    \item Miras with an unknown spectral type (Fig.~\ref{Fig:Spectra 1}):
    Table~\ref{Tab:C-type candidate Miras} lists 20 stars with no previous spectral type determination that we suggest are new carbon stars. Fourteen of them have a BP/RP spectrum characteristic of C stars, with $W_{RP}-W_{K}>0.6$, $mdwlrp>7$, and in most cases $(J-K_{\rm S})>2$. The same holds for another four of them, but a 'C-rich?' classification has been assigned to the BP/RP spectrum because they are noisier or have noticeable weak peaks that could be related to an O-rich nature, so an unambiguous classification in terms of their BP/RP spectrum is not possible. In addition, ASAS J072838-4705.2 does not have an \texttt{mdwlrp} value and ASAS J191421-0519.1 has $mdwlrp<7$ (isCstar=0), although both appear as C-rich in their BP/RP spectra and their $W_{RP}-W_{K}$ colours are typical of C-rich stars.
    
    \item Miras with previous spectral type assignment (Fig.~\ref{Fig:Spectra 1}):
        A possible change in spectral classification is suggested for five Miras, as all of them have a BP/RP spectrum and a $W_{RP}-W_{K}$ value that place them in the C-rich regime. ASAS J082600-6331.5 has a noisy spectrum but with C-rich characteristics. EW~Pup is a Mira of previous spectral type Se with an unambiguous BP/RP spectrum and a value of $mdwlrp>7$. ASAS J183136-1815.4 has a previous spectral type Me, although it has a value of $mdwlrp < 7$ and could be a false positive. ASAS J083508-0944.0 has a previous spectral type M5/6 and a value of $mdwlrp > 7$. WY~Cam is a Mira with a previous spectral type S2e with clear band heads due to CN absorption and also has $mdwlrp > 7$.

        On the other hand, VX~Aql is a transition-type object with spectral type SC5.8. Its spectrum shows characteristics typical of stars with ${\rm C}/{\rm O}\approx1$, even at $mdwlrp>7$, which justifies its current classification. Transition-type objects such as these can change their spectral type on a relatively short timescale, and their periods can be intrinsically unstable, making them particularly interesting study objects \citep{Zijlstra2004,Uttenthaler2016b}.
        
\end{itemize}

\section{Discussion and conclusions}\label{sec: discussion and conclusions}

We collected a large sample of 3100 Mira variables from the ASAS and ASAS-SN databases that cover several pulsation cycles of the stars (Sect.~\ref{sec:Sample stars and data reduction}). We inspected the shapes of the light curves and distinguished between ones that have symmetric (S) or asymmetric (A) curves. This information was combined with the spectral type determination, where available. Based on this, we carried out a population analysis of the various groups of Miras, mostly comparing the oxygen-rich, M-type Miras without light-curve bumps (M(NB)) to the ones with bumps (M(B)) and the S and C Miras.

In the first step, we showed that M(NB) Miras have, on average, shorter periods than the M(B) Miras. The period distribution of the M(NB) stars peaks at $\sim220$\,d, whereas that of the M(B) peaks at $\sim330$\,d with very few stars at $P<250$\,d (Fig.~\ref{Fig:Histogram M - type}). Considering that the pulsation period is a function of the initial stellar mass, this may already indicate a difference in mass between the two groups.

The LFs of the Mira groups were then studied by adopting the absolute $K$ magnitude, $M_{K,0}$, as a proxy of luminosity (Sect.~\ref{subsect:LFs}). The selection of stars was restricted here to the ones with more reliable parallaxes; that is, $\varpi/\sigma_{\varpi}>4.0$. As may already be expected from the period distribution, the M(NB) Miras are the faintest group, fainter than the other groups by $\sim0\fm8$. The M(B), S, and C Miras, on the other hand, differ little in their LFs. In Sect.~\ref{subsect:LFs}, we demonstrated that the Tc-rich M- and MS-type Miras have a very high probability of having the same $M_{K,0}$ distribution as the one of M-type Miras with light-curve bumps, suggesting that they originate from the same parent distribution. This agrees well with the result from Paper~I, in which we showed that many M(B) Miras are indeed Tc-rich or have enhanced $^{12}$C/$^{13}$C, both of which indicate 3DUP activity. Based on this connection between light-curve asymmetries and 3DUP activities, a lower limit of $P\approx250$\,d emerges for the occurrence of 3DUP events.

For the same selection of stars with $\varpi/\sigma_{\varpi}>4.0$, we also constructed a PL diagram and made linear fits to the resulting relations. We find that the longer-period M(B) stars have a markedly steeper PL relation than the M(NB) Miras; it is also steeper than the ones of the S and C Miras. The steeper slope may be caused by some of the long-period M(B) stars undergoing hot bottom burning that produces extra flux, on top of the nominal core-mass-luminosity relation.

Section~\ref{subsect: galactic distribution} demonstrated that the distribution of distances to the Galactic midplane also shows a strong similarity between M(B), S- and C-type Miras. We find scale heights of $z_0=209$, 227, and 206\,pc for the M(B), S, and C Miras, respectively. The progenitor masses can be estimated from these scale heights, which gives a range of $1.5M_{\sun}\lesssim M \lesssim2.5M_{\sun}$, consistent with the typical masses of the Miras in the TP-AGB phase. In contrast, the scale height of the M(NB) Miras is $z_0=506$\,pc, indicating progenitors with masses $\lesssim1.5M_{\sun}$. Those with periods $<200$\,d will have masses $<1M_{\sun}$ \citep{Feast2009} and are thus expected to never undergo 3DUP.

We find significant differences between the Miras of Groups~A and S concerning the dust mass-loss rates as measured by the $K-[22]$ index as a function of the pulsation period (Fig.~\ref{Fig:K-[22] SpT}). The linear relation given by Eq.~\ref{Eq: BSL K-[22]} separates Groups~A and S with an accuracy of 0.89 and exactly matches the one found by \citet{Uttenthaler2019} for the line that best separates Tc-rich and Tc-poor Miras. Similar trends are observed in the $[3.4]-[22]$ versus period diagram (Figs.~\ref{Fig:W1-W4 SpT} and~\ref{Fig:W1-W4 vs period}), though saturation is an issue in the $[3.4]$ band. The vast majority of the M(NB) Miras are located in the region where Tc-poor Miras preferentially are located, and most of the M(B) Miras are located in the zone with lower $K-[22]$ or $[3.4]-[22]$ where Tc-rich S- and C-type post-3DUP Miras preferentially are located. This evidence reinforces the hypothesis that the presence of asymmetries in the light curves is closely linked to 3DUP processes; that is, they correspond to the same evolutionary stage on the TP-AGB. This would allow us to place, with more than 90\% confidence, an M-type Mira in different evolutionary stages based only on the presence or absence of asymmetries.

A non-negligible fraction of the S-type Miras in the sample ($\sim 12\%$) has $K-[22]<0.85$, a level at which no significant dust-mass loss might occur \citep{McDonald2016}. In Paper~I, we suggested that reduced dust formation in the stellar envelope due to effects induced by 3DUP could affect the shape of the light curve. We also refer to \citet{2024A&A...690A.393U} for observational evidence on the effect of 3DUP on dust formation. If dust forms inefficiently in the wake of an outward-propagating shock wave, or is less effectively coupled to the radiation pressure of the central star, significant amounts of material would fall back towards the star instead of being accelerated outwards. This material would collide with the outward-travelling shock wave in the next pulsation cycle, which could cause bumps or secondary maxima in the light curve. Less dust would also mean a reduction in the mid-IR colour, exactly what is found for Tc-rich post-3DUP Miras. Radial-velocity monitoring of lines originating in the dust-forming layers of Miras with light-curve bumps would be an important observational test of this hypothesis. \citet{KR1994} suggest that the duration of a light curve bump equals the time it takes for the shock wave to pass through the stellar atmosphere. They find that the light curve bumps have very similar durations, of the order of a tenth of the period. In this picture, stars with bumps may simply be those with more extended atmospheres, in which the shock wave takes a longer time to travel to the uppermost photospheric layers.

The possible population splitting described above should place the different groups of Miras in the sample in different regions of a {\it Gaia}-2MASS diagram, which we display in Fig.~\ref{Fig:WRP-WK vs MK SpT}. Almost all of the M-type Miras in the sample are found in the O-rich low-mass AGB zone of stars in the initial mass range of $0.9\lesssim M_{i}/M_{\sun}\lesssim1.8$. The distribution of the M(NB) stars clearly differs from that of the M(B) and S Miras, with the latter occupying regions corresponding to a mass range of $1.4\lesssim M_{i}/M_{\sun}\lesssim1.8$, but less evolved than the carbon stars \citep{Lebzelter2018}.

In the zone of O-rich intermediate-mass AGB ($2M_{\sun}\lesssim M_{i}\lesssim3.2M_{\sun}$), we find about 20 stars, mostly M(B) Miras clustered in the faint part of this branch and a few S Miras. These will move quickly to the C-rich branch as soon as they undergo additional 3DUP events. We also find some M(B) Miras in the brightest part of this branch populated by intermediate-mass stars in the range $3.2\lesssim M_{i}/M_{\sun}\lesssim6.0$ that will probably undergo HBB and never evolve to the C-rich region. However, we must acknowledge here that variability-induced photometric scatter and the fact that the {\it Gaia} and 2MASS observations were taken some 15-20 years apart might move a few stars to branches in that diagram where they actually do not belong.

We also observe from Fig.~\ref{Fig:WRP-WK vs MK SpT} that all C-rich Miras are located on the right-hand side of the diagram. They are separated from the O-rich stars by a gap with low stellar density in the range $-0.5\lesssim W_{RP}-W_{K}\lesssim0.6$. Stars are expected to rapidly transit between the O-rich to the C-rich regime as soon as repeated 3DUP episodes lift the C/O ratio above unity.

Figure~\ref{Fig:WRP-WK vs MK NO SpT} in Appendix~\ref{Appendix A} shows the same study for Group~A and S stars of unknown spectral type. Again, Group~A stars are clustered towards the more luminous part of the O-rich low-mass AGB zone, while Group~S stars are shifted towards lower luminosities. Objects in the C-rich and Extreme C-rich zones have been studied for possible classification as C-type stars (Sect.~\ref{sec: New carbon stars}).

We also propose a new parameter that we call the distance to the best separation line (DBSL), defined as the difference between the observed colour index of a Mira from the line that best separates the Group~A and S Miras in that colour index versus period $P$ (Eq.~\ref{Eq: BSL K-[22]} and Eq.~\ref{Eq: BSL W1-W4}). The colour indices are $K-[22]$ or $[3.4]-[22]$. We constructed new diagrams of the DBSL parameters as a function of the Wesenheit index difference. The sample Miras occupy the four quadrants of this diagram according to their evolutionary state. Beginning with the upper left quadrant (Q1), we find the M-type Miras without light-curve bumps, which are mostly Tc-poor and the least evolved sample stars in terms of progression of the 3DUP state (M(NB)). Moving in an anticlockwise direction, the Q2 quadrant on the lower left contains M-type Miras, typically with bumps, and S-type Miras. Once a star has experienced sufficient 3DUP events to raise C/O above unity, the star will move to the lower right into the C-rich regime. Finally, among the carbon stars, the DBSL parameter (and thus $K-[22]$ or $[3.4]-[22]$) and the Wesenheit index difference will progressively increase, and the star will ultimately shift to the extreme C-rich zone in quadrant Q4 on the upper right of the diagram.

The evidence gathered in this research establishes a dichotomy within the M-type Miras, placing the M(B) Miras in the early stages of the TP-AGB phase, just after the first TPs. They would represent the link between M-type and S-type Miras, embedded in the evolutionary sequence M(NB) -- M(B) -- S -- SC -- C in the area occupied by MS-type Miras. They will evolve into S- and C-type Miras as TPs and 3DUP processes occur, gradually enriching their atmospheres with processed material from the interior and progressively increasing the C/O ratio. The fact that they form a relatively homogeneous group makes them a testbed for studying the changes induced by the first TPs in the internal structure of the Miras, in dust formation and in the stellar winds that determine the mass losses in these final evolutionary stages. The existence of data covering several pulsation cycles may be sufficient to place them in this phase with a degree of confidence above 90\%.

\section{Data availability}\label{sec:Data availability}

The data resulting from the analysis in this paper are only available in electronic form at the CDS via anonymous ftp to \url{cdsarc.u-strasbg.fr} (130.79.128.5) or via \url{http://cdsweb.u-strasbg.fr/cgi-bin/qcat?J/A+A/}.

\begin{acknowledgements}

The authors thank Miguel Jurado Vargas for discussions and advice on the manuscript, the referee for constructive comments that helped improve the manuscript, and George Gontcharov for kindly providing the 2023 extinction map. This research was funded in part by the Austrian Science Fund (FWF) 10.55776/F81. For open access purposes, the authors have applied a CC BY public copyright license to any author accepted manuscript version arising from this submission. This work has made use of data from the European Space Agency (ESA) mission {\it Gaia} (\url{https://www.cosmos.esa.int/gaia}), processed by the {\it Gaia} Data Processing and Analysis Consortium (DPAC, \url{https://www.cosmos.esa.int/web/gaia/dpac/consortium}). Funding for the DPAC has been provided by national institutions, in particular the institutions participating in the {\it Gaia} Multilateral Agreement. This publication makes use of data products from the Two Micron All Sky Survey, which is a joint project of the University of Massachusetts and the Infrared Processing and Analysis Center/California Institute of Technology, funded by the National Aeronautics and Space Administration and the National Science Foundation. This publication makes use of data products from the Wide-field Infrared Survey Explorer, which is a joint project of the University of California, Los Angeles, and the Jet Propulsion Laboratory/California Institute of Technology, funded by the National Aeronautics and Space Administration.
\end{acknowledgements}

\bibliographystyle{aa}
\bibliography{aa53622-24}

\begin{appendix}

\section{Analysis of Mira-type variables with an unknown spectral type}\label{Appendix A}

We present the analysis of the 1338 stars with unknown spectral types of the same properties that we investigated in the Miras with known spectral types throughout the project. The distribution of these stars in terms of their variability type was shown in Table~\ref{Tab:sample}. Here the stars are not divided on the basis of their spectral types, but only taking into account the presence or absence of asymmetries in their light curves. Therefore, the results could reinforce the hypotheses put forward in this project independently of atmospheric chemistry. 

\begin{figure}
\includegraphics[width=\linewidth]{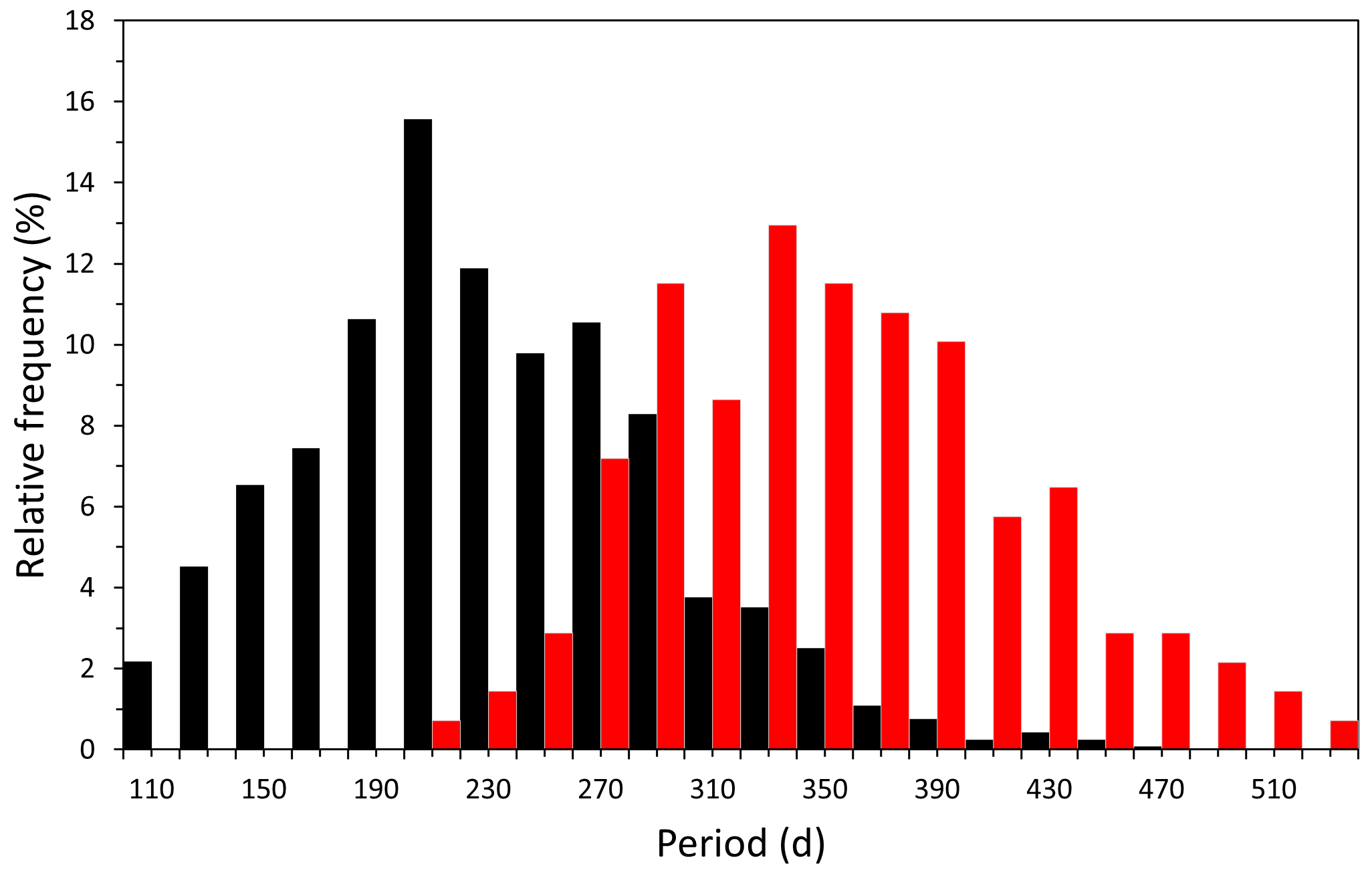}
\centering
\caption{Histogram of the pulsation periods of stars with an unknown spectral type, with intervals of 40 days.}
\label{Fig:Hist no SpT}
\end{figure}

We treated the light curves of the stars following the same criteria described in Sect.~\ref{sec:Sample stars and data reduction}, then separated them into the two Groups~S and A according to the absence or presence of light-curve asymmetries. We found 1198 stars belonging to Group~S ($\sim 89.5\%$) and only 140 in Group~A ($\sim10.5\%$). These fractions are quite different from those found in the Miras with a known spectral type, probably because they are on the whole less luminous, and therefore have shorter pulsation periods and a higher probability of no asymmetries in their light curves. Nevertheless, the period distribution follows a similar trend to that found for the Miras with a known spectral type, as we can corroborate in Fig.~\ref{Fig:Hist no SpT}. Group~S and A stars have average periods of $\left<P\right>=231$\,d and $353$\,d, respectively.

\begin{figure*}
\includegraphics[scale=0.55]{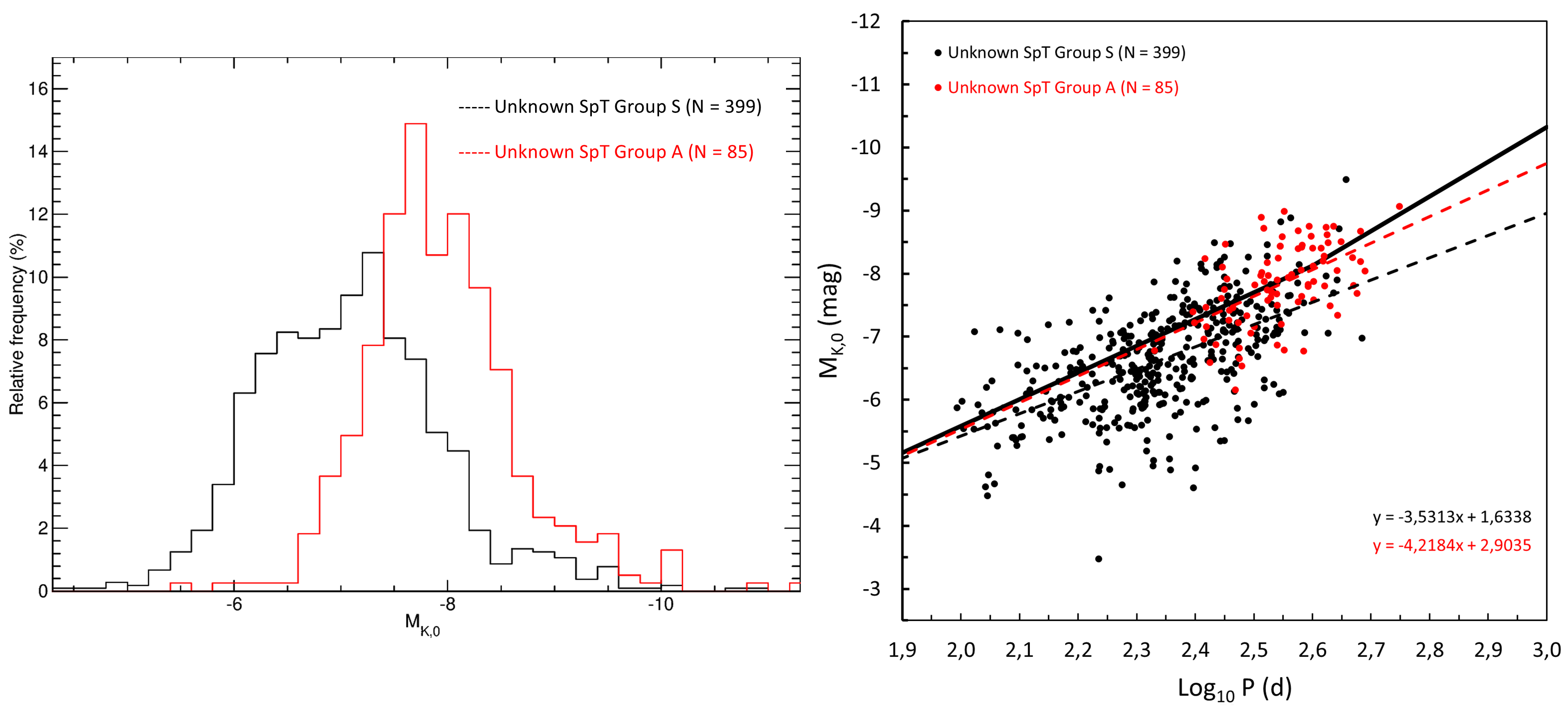}
\centering
\caption{$M_{K,0}$ histrogram and P-L-relations for Miras with an unknown spectral type in the sample. The thick black line in the PL diagram on the right-hand side indicates the relation obtained by \citet{Sanders2023b} for Galactic Miras.}
\label{Fig: MK - period NO SPT}
\end{figure*}

In the left-hand panel of Fig.~\ref{Fig: MK - period NO SPT}, the LFs of the stars in Groups~A and S are plotted. A similar trend to the one discussed above for the Miras of known spectral type can be observed, with different patterns in the LFs of the stars in Groups S and A. The Group~S stars have mean absolute magnitudes of $\langle M_{K,0}\rangle=-6.63\pm0.84$, compared to $\langle M_{K,0}\rangle=-7.83\pm0.64$ for the Group~A stars. The PL diagram for these stars, shown in the right panel of Fig.~\ref{Fig: MK - period NO SPT}, shows a similar trend. The linear fits made for these groups of stars are shown in Table~\ref{Tab:P-L no SpT}, although they should be treated with caution due to the significant luminosity bias on stars of unknown spectral type discussed above. In the data from these fits we can see that the Group~A stars have a steeper slope than the Group~S stars, consistent with the results in the main part of this paper, where from a threshold period of $P\approx400$\,d there is a significant increase towards higher luminosities.

\begin{table}
\caption{Linear P-L fits to the different groups of stars with unknown spectral types in our sample.}
\label{Tab:P-L no SpT}
\centering
\begin{tabular}[b]{l c c c}
\hline \hline
Group & slope ($\rho$) & Intercept & Zero point ($\delta$)\\
\hline
All stars & -4.18 $\pm{0.20}$ & 3.08 $\pm{0.49}$ & -6.86 \\
Group~S & -3.53 $\pm{0.25}$ & 1.63 $\pm{0.59}$ & -6.77 \\
Group~A & -4.22 $\pm{0.73}$ & 2.90 $\pm{1.87}$ & -7.14 \\
\hline
\end{tabular}
\end{table}

\begin{figure*}[t]
\includegraphics[scale = 0.6]{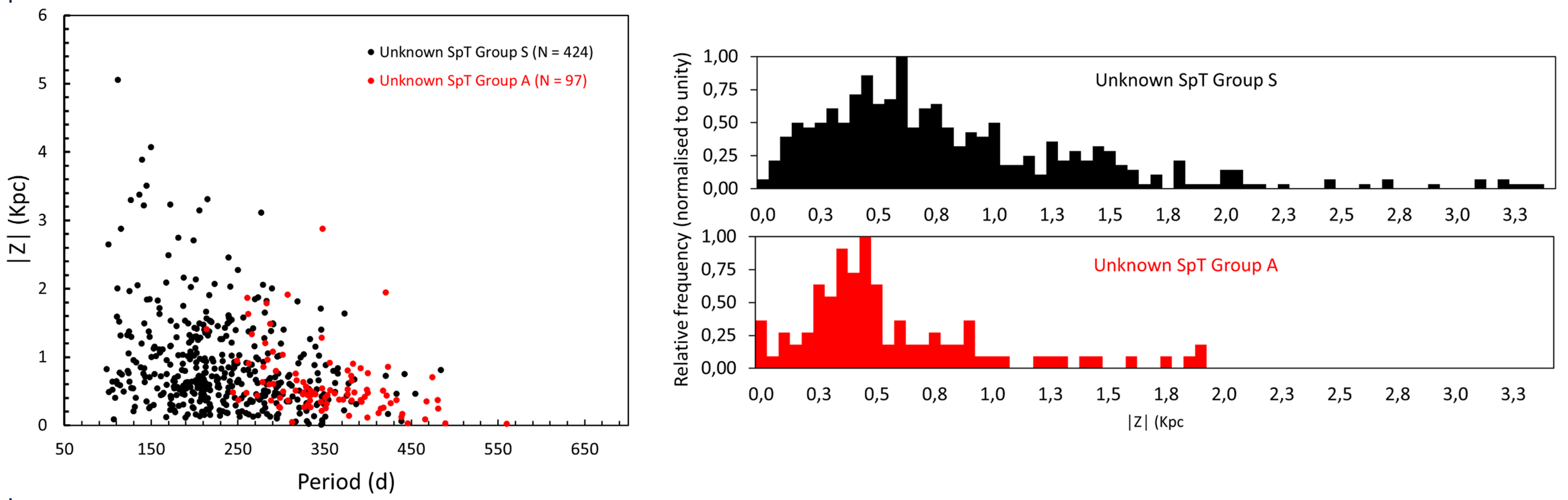}
\centering
\caption{\textit{Left panel}: Distance to the Galactic midplane |Z| as a function of the pulsation period P. \textit{Right panel}: Histogram of the relative star frequency as a function of |Z| for the two groups of stars with an unknown spectral type.}
\label{Fig:Z Midplane NO SpT}
\end{figure*}

The Galactic distribution of the two groups of stars also shows significant differences. In the left-hand panel of Fig.~\ref{Fig:Z Midplane NO SpT}, the distance to the Galactic midplane $|Z|$ is plotted versus the pulsation period $P$. As we already observed in Sect.~\ref{subsect: galactic distribution} for the Miras with a known spectral type (see Fig.~\ref{Fig:Z Midplane} and Fig.~\ref{Fig:Z Midplane 2}), it is remarkable that both the value of $|Z|$ and the upper envelope of the $|Z|$ distribution show a clear decrease with increasing pulsation period, reflecting the period-age relation of Miras \citep{Trabucchi2022}. However, Group~A stars are mostly distributed at $\langle|Z|\rangle\lesssim1$\,kpc, with a few of them in the range $1\lesssim\langle|Z|\rangle\lesssim2$\,kpc, while Group~S stars are homogeneously distributed up to values of $\langle|Z|\rangle\approx4$\,kpc. In Table~\ref{Tab:Galactic distribution no SpT}, we see that the stars in Groups S and A have a mean value of the distance to the Galactic midplane of $\langle|Z|\rangle=0.88$\,kpc and $\langle|Z|\rangle=0.61$\,kpc, respectively, with clear differences between the two groups and in agreement with the results obtained from the Miras with a known spectral type.

\begin{table} 
\caption{Galactic distribution of stars with unknown spectral types.}
\label{Tab:Galactic distribution no SpT}
\centering
\begin{tabular}[b]{c c c c c c}
\hline \hline
Type & $\langle|Z|\rangle$ & $|Z|$            & $|Z|$            & $1/z_{0}$    & R\\
     & (kpc)               & 90$^{\rm th}$ p. & 95$^{\rm th}$ p. & (kpc$^{-1}$) &  \\
\hline
Group~S & 0.88 & 1.61 & 2.06 & 1.88 & 0.59\\
Group~A & 0.61 & 1.08 & 1.41 & 3.57 & 0.24\\
\hline
\end{tabular}
\tablefoot{Column~1: Group designation; column~2: mean distance to the Galactic midplane; columns~3 and 4: 90$^{\rm th}$ and 95$^{\rm th}$ percentiles of $|Z|$; column~5: distance scale; column~6: goodness of fit $R$ to the histograms in the right panel of Fig.~\ref{Fig:Z Midplane NO SpT}.}
\end{table}

\begin{figure*}[t]
\includegraphics[scale = 0.7]{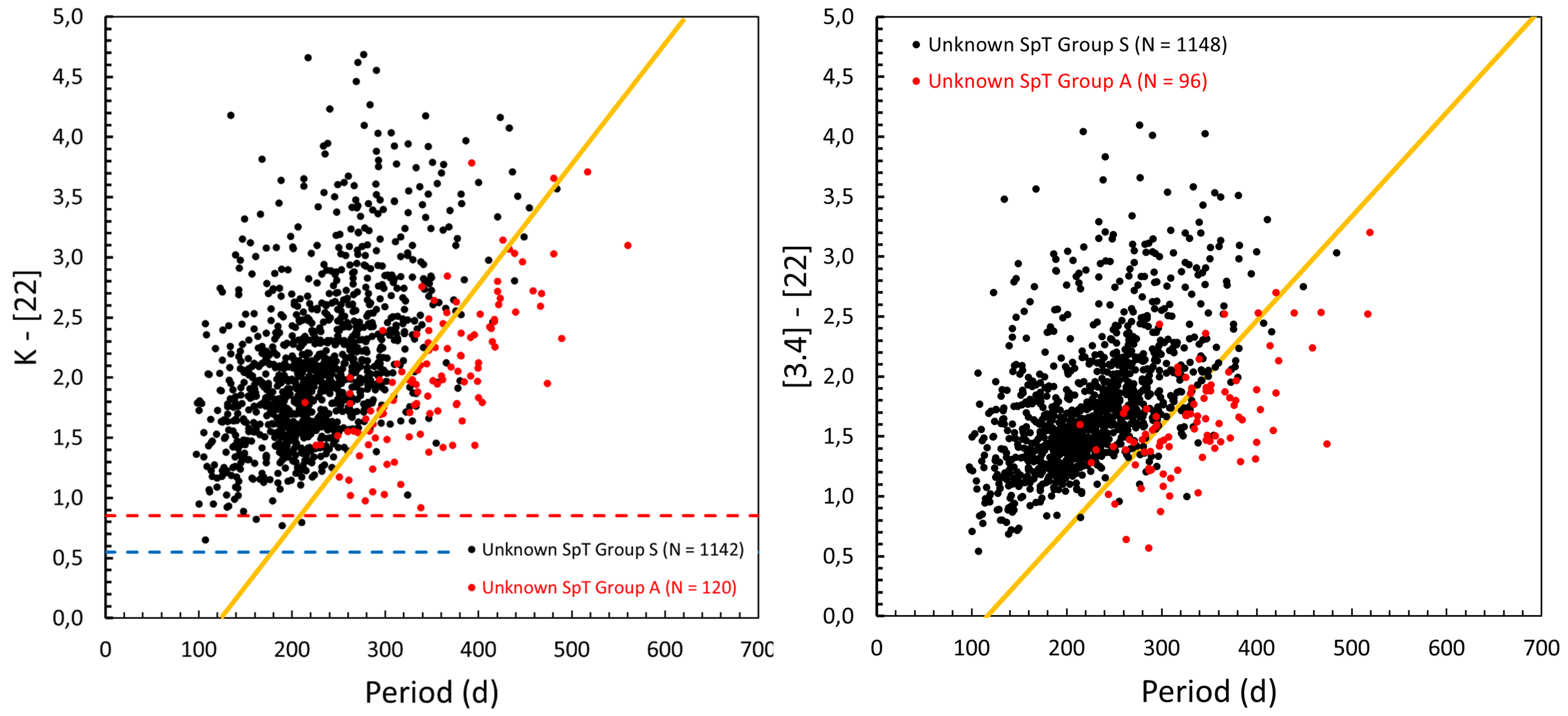}
\centering
\caption{$K-[22]$ vs\ P (left panel) and $[3.4]-[22]$ vs\ P (right panel) diagrams for the Miras with an unknown spectral type of the sample. See the text for the description of these diagrams.}
\label{Fig:CD NO SpT}
\end{figure*}

\begin{figure*}[!t]
\includegraphics[width=0.5\linewidth]{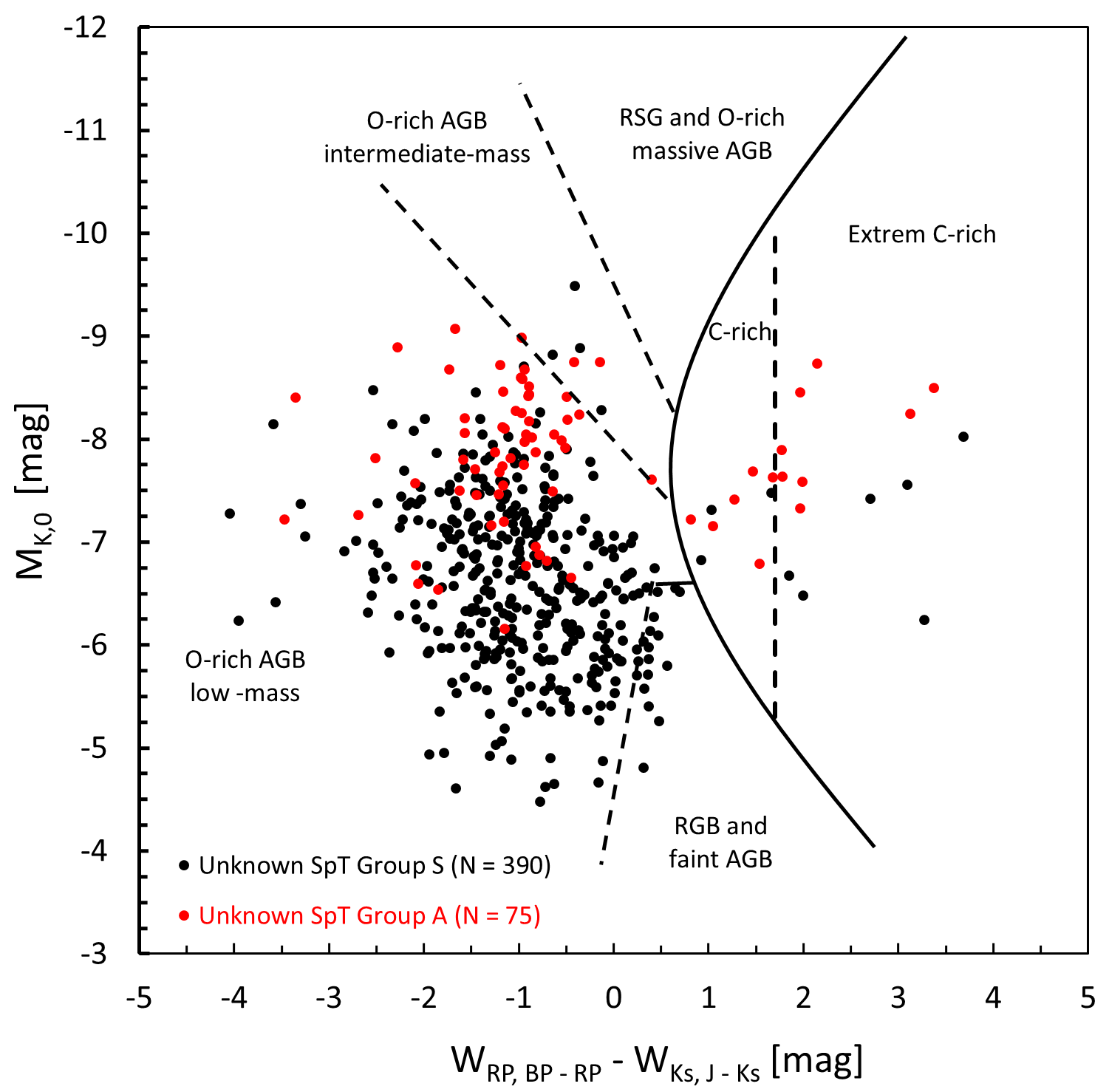}
\centering
\caption{$W_{RP,RP-BP}$ vs\ $M_{K,0}$ for stars with an unknown spectral type. See legends to identify the Mira type and the number of stars in each group.}
\label{Fig:WRP-WK vs MK NO SpT}
\end{figure*}

The right panel of Fig.~\ref{Fig:Z Midplane NO SpT} plots the histograms of both groups of stars, where the relative frequency, normalised to unity, is plotted as a function of $|Z|$. There is a clear difference between Groups~S and A. Table~\ref{Tab:Galactic distribution no SpT} shows the results of fitting an exponential of the form $N = N_{0} e^{-|Z|/z_{0}}$ to these histograms, yielding a value of $z_{0}=280$\,pc for Group~A, which, although slightly higher than the range of $150-250$\,pc estimated by \citet{Claussen1987} for the progenitors of the C-type Miras, is fairly consistent with this range. For the Group~S stars, we obtain a value of $z_{0}=532$\,pc, indicative of progenitors with significantly lower masses. It is noteworthy to mention that the goodness-of-fit parameter is much lower than for the stars with known spectral types (column 6 of Table~\ref{Tab:Galactic distribution no SpT}). While this may be related to the small number of stars in Group~A, also the dearth of stars at small distances to the Galactic midplane plays an important role. This is due to a selection effect because surveys to determine spectral types of stars may have focused on low Galactic latitudes and may have neglected stars at higher latitudes. Indeed, an Aitoff projection of the stars with unknown spectral types shows a 'zone of avoidance' at low Galactic latitudes. Other than that, the results for the stars with unknown spectral types agree with those for the stars with known spectral types in the main part of the paper (Sect.~\ref{subsect: galactic distribution}).

Analysing the mass-loss rates for Miras belonging to Groups~S and A, it is very interesting to check that they follow a similar trend to the one found in Sect.~\ref{sec:dust-MLR} for the Miras with a known spectral type, further reinforcing the possible connection between the asymmetries in the light curves and the dust mass-loss rates, regardless of the knowledge of their spectral type. In the left panel of Fig.~\ref{Fig:CD NO SpT}, we plot the $K-[22]$ versus $P$ diagram for the 1262 Miras with an unknown spectral type and no saturation in either the $K$ or [22] bands. Analogous to  Fig.~\ref{Fig:K-[22] SpT}, the solid orange line is the relation given by Eq.~\ref{Eq: BSL K-[22]}, the dashed red line corresponds to the limit of $K-[22] = 0.85$ \citep{McDonald2016},  and the dashed blue line corresponds to the limit of $K-[22] = 0.55$ \citep{McDonald2018}. We can appreciate that the Group~A Miras are preferentially located below the orange separation line, while the Group~S Miras are located above this line. Figure~\ref{Fig:K-[22] vs Period} shows in more detail the distribution of Group~S and A Miras and their fractions above and below the separating line given by Eq.~\ref{Eq: BSL K-[22]}, respectively.

In the right-hand panel of Fig.~\ref{Fig:CD NO SpT}, we see how these two groups are distributed in the $[3.4]-[22]$ versus $P$ diagram, with 1244 unsaturated Miras in the [3.4] and [22] bands. The distribution is similar to that in the $K-[22]$ versus $P$ diagram. A fraction of $\sim 96.8\%$ of the Group~S Miras lie above the dividing line given by Eq.~\ref{Eq: BSL W1-W4}, while those of Group~A lie below it by a fraction of $\sim 72.9\%$. This slightly lower percentage with respect to the Miras of known spectral type is probably due to the bias mentioned in Sect.~\ref{sec:dust-MLR}, caused by the higher saturation in the [3.4] and [22] bands.

Fig.~\ref{Fig:WRP-WK vs MK NO SpT} plots the {\it Gaia}-2MASS diagram for stars with an unknown spectral type, where we can see a similar distribution among the Miras of Groups A and S to those found for the stars with a known spectral type (see Sect.~\ref{sec: Gaia-2MASS photometry}). Most of the Group~A Miras are located in the bright part of the O-rich low-mass AGB zone, between $-9.0\lesssim M_{K,0}\lesssim-6.8$, while the Group~S Miras are shifted to higher $M_{K,0}$ values, as can be inferred from the LFs shown in Fig.~\ref{Fig: MK - period NO SPT}. It can be seen how some of them are even making an incursion into the ‘RGB and faint AGB’ zone, possibly still in pre-AGB stages. On the other hand, only six Miras are found in the O-rich intermediate-mass AGB zone, while the O-rich massive AGB and RSG zone is completely unpopulated, as could be expected from the LFs discussed above. Also, note the clear gap observed between the C-rich zone and the rest, similar to the one observed in Fig.~\ref{Fig:WRP-WK vs MK SpT} for the Miras with a known spectral type. Twenty-three Miras are located in the C-rich and Extreme C-rich region of Fig.~\ref{Fig:WRP-WK vs MK NO SpT}, 20 of which have been proposed as new carbon stars (Sect.~\ref{sec: New carbon stars}), and are listed in Table~\ref{Tab:C-type candidate Miras}.

\begin{figure*}[t]
\includegraphics[width=\linewidth]{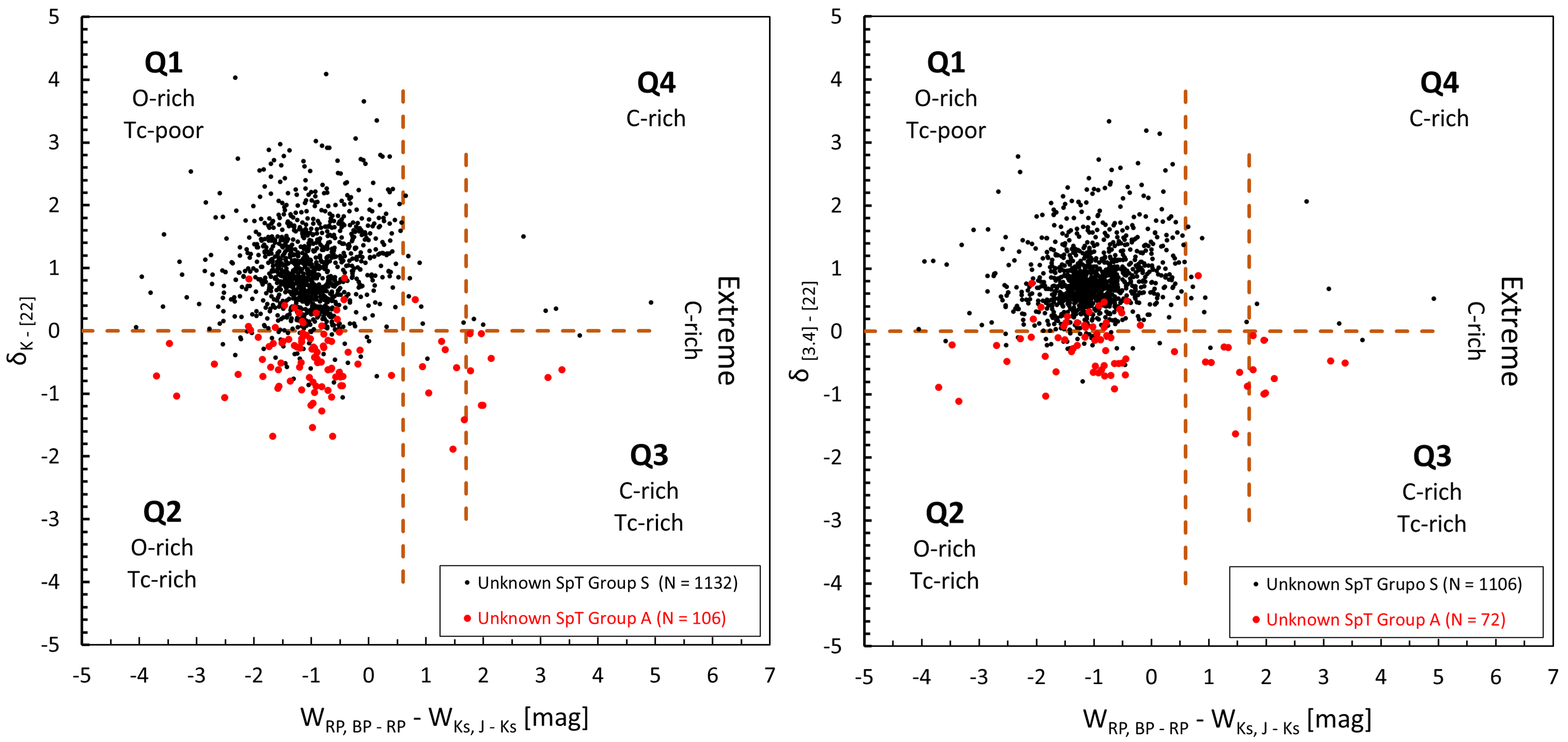}
\centering
\caption{\textit{Left panel:} $\delta_{K-[22]}$ vs\ $W_{RP} - W_{K}$ diagram. \textit{Right panel:} $\delta_{[3.4]-[22]}$ vs\ $W_{RP} - W_{K}$ diagram. See legends to identify the Mira type and the number of stars in each group.}
\label{Fig:DBSL1 NO SPT}
\end{figure*}

\begin{figure*}[h]
\includegraphics[scale=0.7]{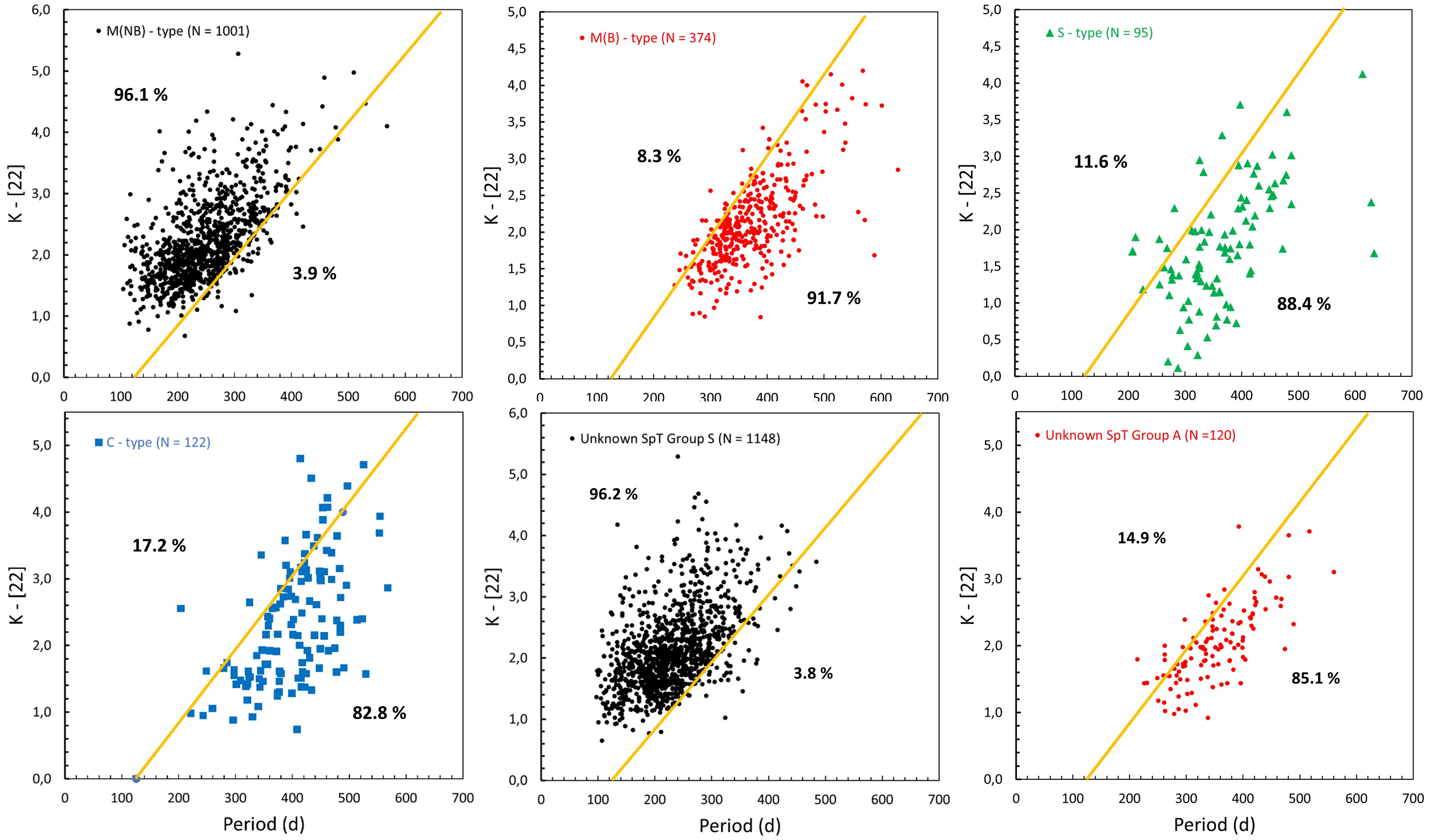}
\centering
\caption{$K-[22]$ vs\ period diagrams for the different groups of Mira variables (see legends).}
\label{Fig:K-[22] vs Period}
\end{figure*}

The WISE-2MASS and {\it Gaia}-2MASS photometry features of stars with an unknown spectral type can be compared in the DBSL diagrams introduced in Sect.~\ref{sec: DBSL function}. Fig.~\ref{Fig:DBSL1 NO SPT} shows the DBSL versus $P$ diagrams for the $K-[22]$ (left panel) and $[3.4]-[22]$ (right panel) indices. Although here the number of Miras belonging to Group~A is significantly lower than in the case of Miras with a known spectral type (see Fig.~\ref{Fig:DELTA SPT}), a similar pattern is observed in both diagrams. The first quadrant is almost entirely populated by Group~S stars, as can be inferred from the $K-[22]$ versus P diagram in Fig.~\ref{Fig:CD NO SpT}, while most of the few Group~S stars are in Q2. The twenty or so C-type candidates discussed above are distributed throughout the Q3 and Q4 quadrants, seeming to infer a positive correlation between the DBSL and $W_{RP}-W_{K}$ parameters, similar to that observed in Fig.~\ref{Fig:DELTA SPT}, although here with a more limited number of stars.

The characteristics observed in this short study of stars with unknown spectral types evidently do not take into account the details concerning atmospheric chemistry but exclusively the presence or absence of asymmetries. The fact that the studied features are similar to those observed for Mira with a known spectral type reaffirms the hypotheses put forward in this project.

\onecolumn
\section{Additional tables}

\begin{table*}[!h]
\caption{Stars from ASAS database discarded in the final sample.}
\label{Tab:Discarded stars}
\centering
\begin{tabular}[h]{c c c c}
\hline \hline 
Designation & Var. type & SpT & Cause for discarding\\
\hline 
ASAS  J182452-2903.7 & M & M2 & Composite Object, Blend\\
R Aqr & M &  & Symbiotic Star\\
U Boo & SRB & M4e & Irregular phase diagram\\
V0586 Car & SRB &  & Irregular phase diagram\\
$o$ Cet & M & M5e-M9e & Symbiotic Star\\
CN Cha & LB &  & Irregular phase diagram\\
ASAS J054957-5252.1 & M &  & Irregular phase diagram\\
ASAS J111848-6821.7 & M &  & Irregular phase diagram\\
ASAS J172739-2740.0 & M &  & Irregular phase diagram\\
ASAS J174225-2203.2 & M &  & Irregular phase diagram\\
ASAS J175523-3421.0 & M &  & Irregular phase diagram\\
ASAS J180458-3033.5 & M &  & Irregular phase diagram\\
ASAS J184259-1958.2 & M &  & Composite Object, Blend\\
GH Lib & M &  & Algol, eclipsing systems\\
FI Lup & LB & Me & Irregular phase diagram\\
FS Lyr & L &   & Irregular phase diagram\\
IO Nor & M &  & R CrB Variable\\
V0447 Oph & SRB & M4IIIe & Irregular phase diagram\\
V0851 Oph & SRB & M5 & Irregular phase diagram\\
DH Ori & SRA & M0e & Irregular phase diagram\\
EO Ori & LB & M6 & Irregular phase diagram\\
AM Peg & SRA & M1e-M3e & Irregular phase diagram\\
RY Ret & SRA & M3e & Irregular phase diagram\\
AF Sco & N: &   & Nova\\
V1189 Sco & SRA & Me & Irregular phase diagram\\
WX Sco & IN: &   & Orion variable, eruptive\\
NP Sgr & LB & M5 & Irregular phase diagram\\
W Tau & SRB & M4-M6.5 & Irregular phase diagram\\
\hline
\end{tabular}
\end{table*}

\clearpage

\begin{table*}
\caption{C-type candidate Miras.}
\label{Tab:C-type candidate Miras}
\centering
\begin{tabular}[b]{c c c c c c c c c }
\hline \hline 
Designation & Var. type & SpT & $W_{RP}-W_{K}$ & $J-K_{\rm S}$ & isCstar & BP/RP & New SpT\\
\hline
Known SpT\\
\hline
ASAS J082600-6331.5 & M & M:e & 2.611 & 3.512 & 1 & C-rich? & C-rich\\
EW Pup & M & Se & 1.426 & 1.473 & 1 & C-rich & C-rich\\
ASAS J183136-1815.4 & M & Me & 1.275 & 1.365 & 0 & C-rich & C-rich\\
ASAS J083508-0944.0 & M & M5/6 & 1.274 & 1.391 & 1 & C-rich & C-rich\\
WY Cam & M & S2e & 1.028 & 1.427 & 1 & C-rich & C-rich\\
VX Aql & M & SC5,8 & 0.605 & 2.061 & 1 & C-rich? & C-rich?\\
\hline  
Unknown SpT\\
\hline
ASAS J072838-4705.2 & M & & 3.684 & 2.056 & --- & C-rich & C-rich\\
V0790 Mon & M & & 3.375 & 2.188 & 1 & C-rich & C-rich\\
ASAS J083300-2354.7 & M & & 3.275 & 2.616 & 1 & C-rich & C-rich\\
ASAS J060913-2743.1 & M & & 3.125 & 2.195 & 1 & C-rich & C-rich\\
BD Pyx & SR: & & 3.096 & 1.971 & 1 & C-rich? & C-rich\\
ASAS J155913-2444.1 & M & & 2.141 & 2.726 & 1 & C-rich? & C-rich\\
ASAS J093523-2944.1 & M & & 1.999 & 2.284 & 1 & C-rich? & C-rich\\
V0796 Mon & SR & & 1.991 & 2.009 & 1 & C-rich & C-rich\\
ASAS J074543-4404.3 & M & & 1.964 & 2.240 & 1 & C-rich & C-rich\\
ASAS J182052-4331.9 & M & & 1.963 & 2.362 & 1 & C-rich & C-rich\\
ASAS J191421-0519.1 & M & & 1.844 & 1.805 & 0 & C-rich & C-rich\\
ASAS J084558-2945.1 & M & & 1.776 & 2.200 & 1 & C-rich & C-rich\\
ASAS J080722-5322.4 & M & & 1.772 & 2.301 & 1 & C-rich & C-rich\\
ASAS J184428-1509.3 & M & & 1.674 & 1.803 & 1 & C-rich & C-rich\\
ASAS J192217-1846.1 & M & & 1.660 & 1.919 & 1 & C-rich & C-rich\\
ASAS J051341+1159.2 & M & & 1.537 & 2.213 & 1 & C-rich & C-rich\\
V0379 Hya & SR & & 1,269 & 2.064 & 1 &  C-rich & C-rich\\
ASAS J083432-5401.2 & M & & 1.046 & 1.797 & 1 & C-rich & C-rich\\
BI Ant & SRA & & 1.032 & 1.477 & 1 & C-rich ? & C-rich\\ 
EQ Nor & M & & 0.813 & 1.435 & 1 & C-rich & C-rich\\
\hline
\end{tabular}
\end{table*}

\clearpage

\section{Additional figures}

\begin{figure*}[h]
\includegraphics[scale=0.7]{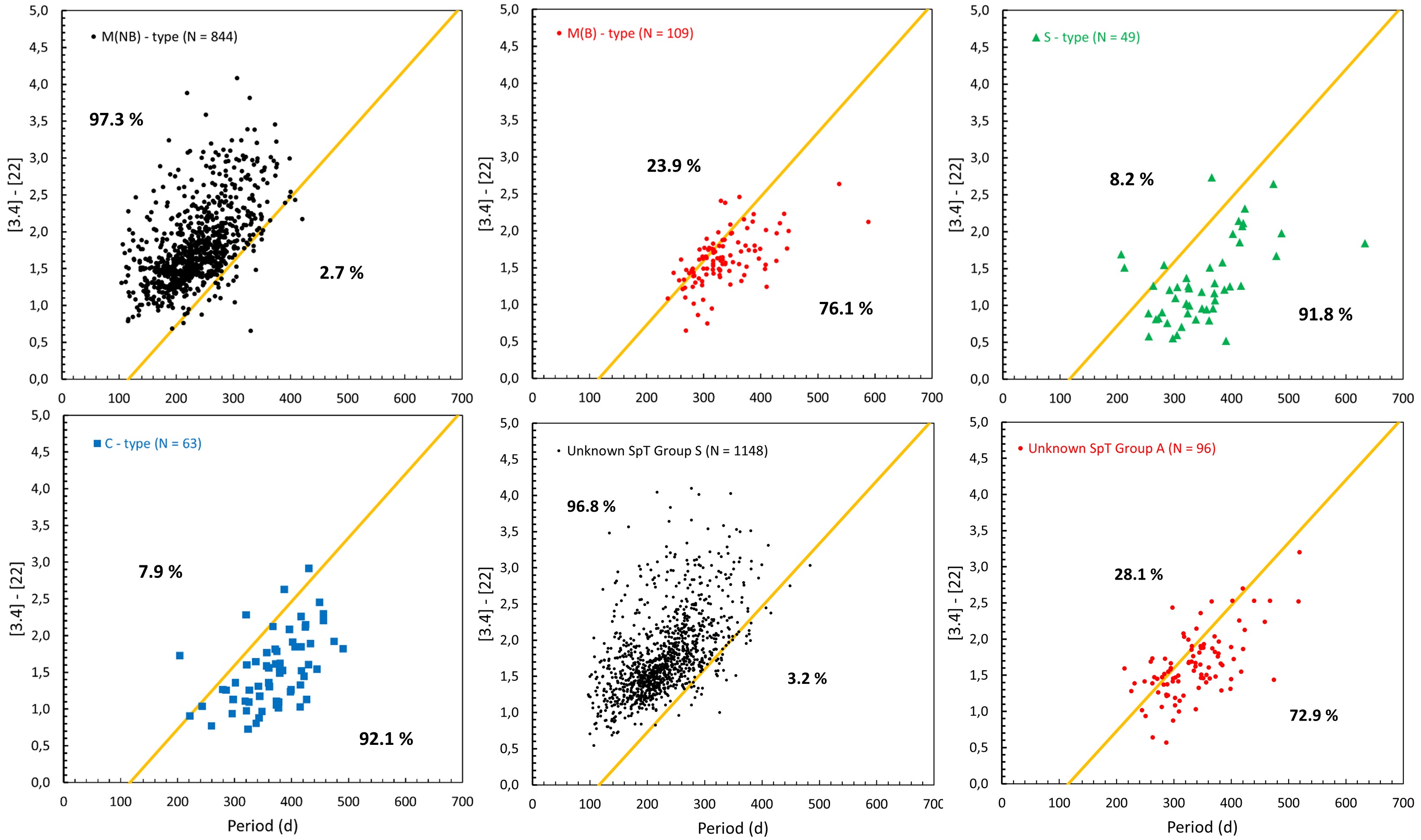}
\centering
\caption{$[3.4]-[22]$ vs\ period diagrams for the different groups of Mira variables (see legends).}
\label{Fig:W1-W4 vs period}
\end{figure*}

\clearpage

\begin{figure}[!t]
\includegraphics[scale=0.45]{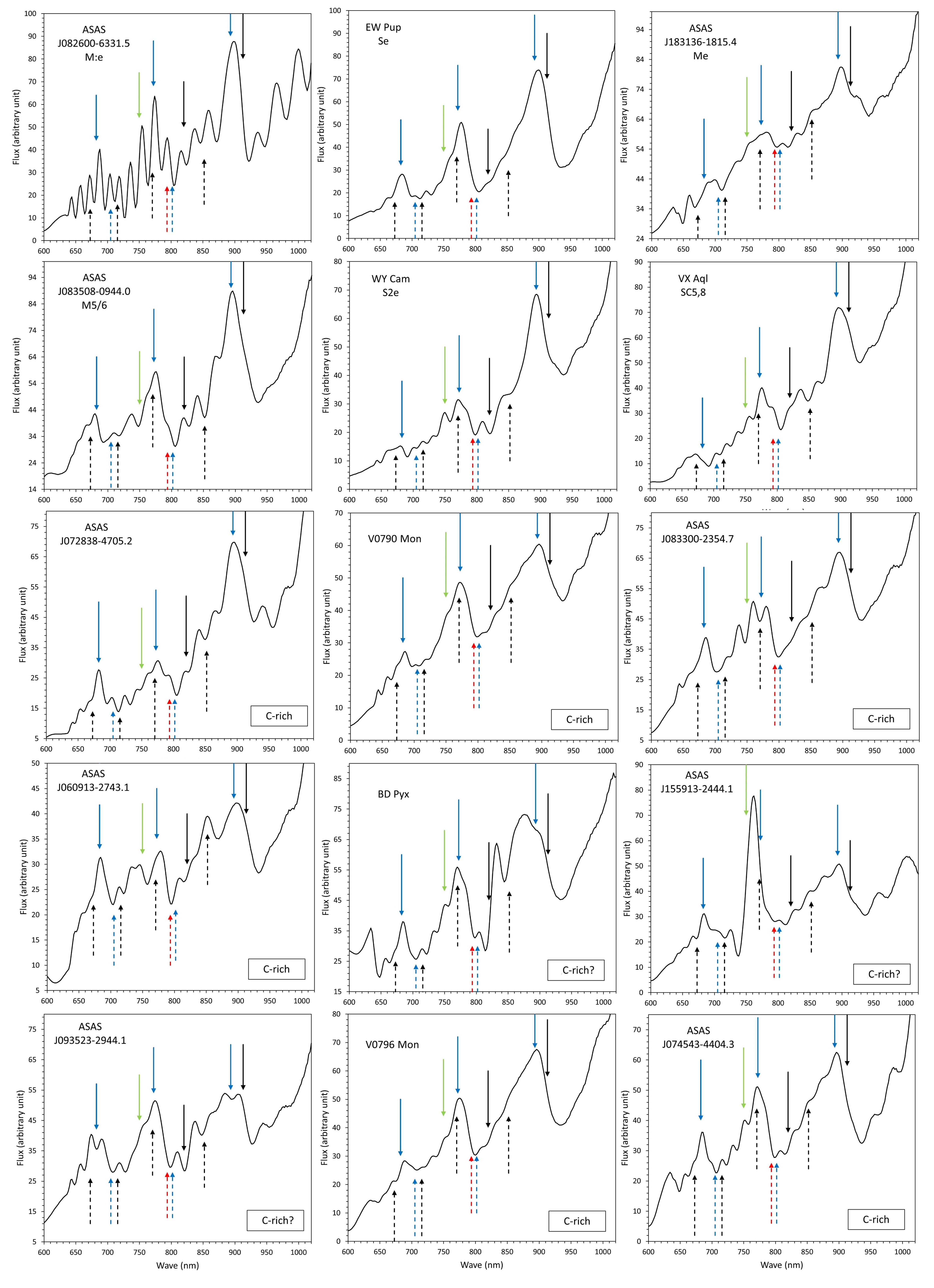}
\centering
\end{figure}

\begin{figure}[!t]
\includegraphics[scale=0.45]{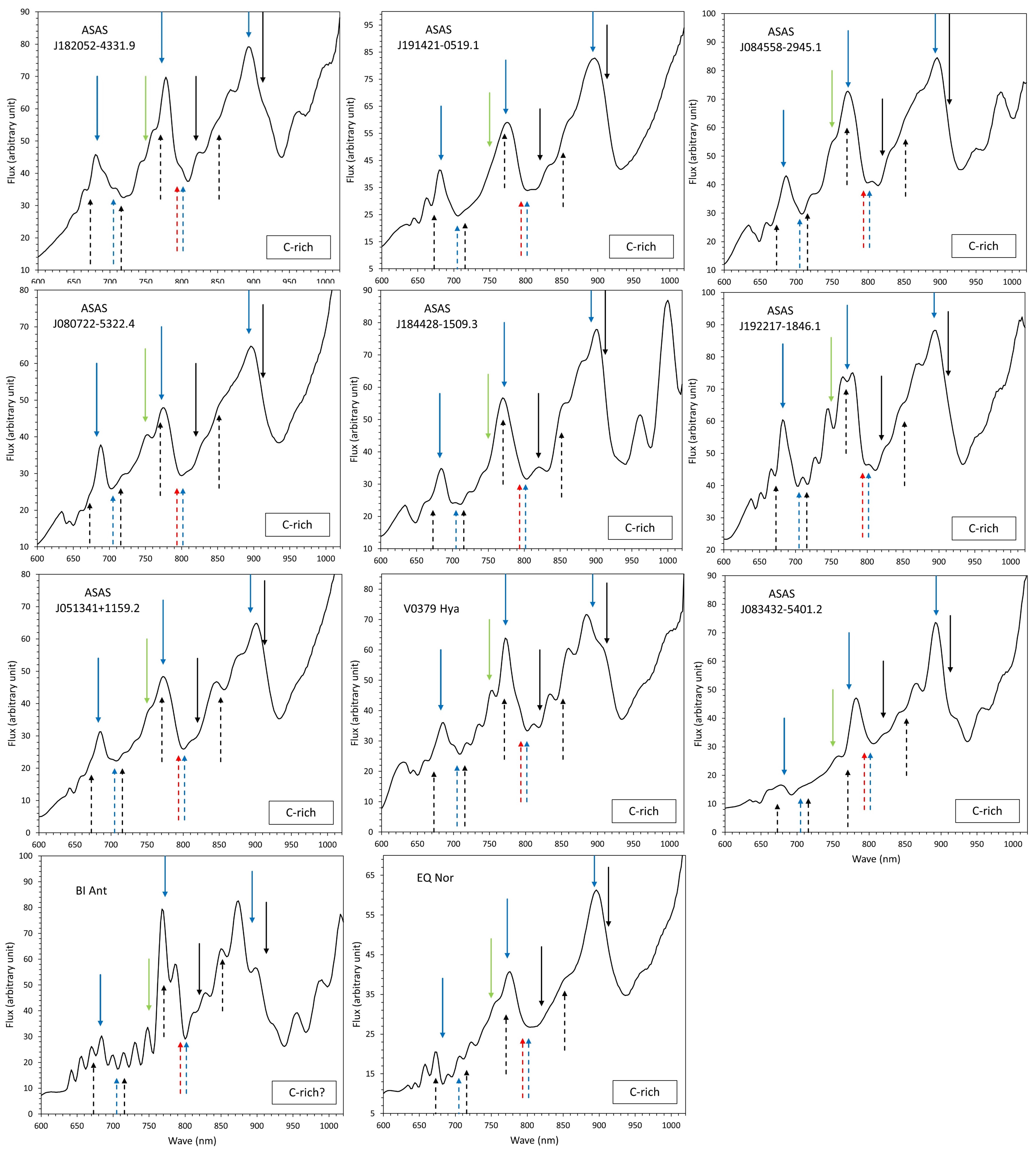}
\centering
\caption{{\it Gaia} BP/RP spectra of Mira variables carbon star candidates included in the table~\ref{Tab:C-type candidate Miras}, in order of appearance, which are located in the C-rich or extremely C-rich zone of Fig~\ref{Fig:WRP-WK vs MK NO SpT}.}
\label{Fig:Spectra 1}
\end{figure}

\end{appendix}

\end{document}